\documentclass[aps,prd,preprintnumbers,superscriptaddress,nofootinbib,onecolumn,notitlepage]{revtex4-1}
\usepackage{ulem}
\usepackage[dvipdfmx]{graphicx}
\usepackage{bm,latexsym,amsmath,amssymb,amsfonts}
\usepackage[colorlinks=true,linkcolor=magenta,citecolor=magenta]{hyperref}
\usepackage{color}

\begin{document}
\title{On Lorentz-invariant spin-2 theories}

\author{Atsushi~Naruko}
\affiliation{Frontier Research Institute for Interdisciplinary Sciences \& 
Department of Physics, Tohoku University \\
Aramaki aza Aoba 6-3, Aoba-ku, Sendai 980-8578, Japan}

\author{Rampei~Kimura}
\affiliation{Waseda Institute for Advanced Study, Waseda University, 1-6-1 Nishi-Waseda, Shinjuku, Tokyo 169-8050, Japan}
\affiliation{Department of Physics, Tokyo Institute of Technology, Tokyo
	152-8551, Japan}

\author{Daisuke~Yamauchi}
\affiliation{Faculty of Engineering, Kanagawa University, Kanagawa-ku,
	Yokohama-shi, Kanagawa, 221-8686, Japan}

\preprint{TU1079, WUAP/1810/18}

\begin{abstract}
We construct Lorentz-invariant massless/massive spin-2 theories in flat spacetime. Starting from the most generic action of a rank-2 symmetric tensor field whose Lagrangian contains up to quadratic in first derivatives of a field, we investigate the possibility of new theories by using the Hamiltonian analysis. By imposing the degeneracy of the kinetic matrix and the existence of subsequent constraints, we classify theories based on the number of degrees of freedom and constraint structures and obtain a wider class of Fierz-Pauli theory as well as massless and partially massless theories, whose scalar and/or vector degrees of freedom are absent. We also discuss the relation between our theories and known massless and massive spin-2 theories.  
\end{abstract}

\maketitle

\section{Introduction} 

The search for a theoretically consistent Lorentz-invariant massive graviton has been a challenging issue since 1939, when Fierz and Pauli (FP) proposed a linear theory of massive spin-2 field \cite{Fierz:1939ix}. 
Once the FP mass term are taken into account in general relativity,
one would naively expect that it recovers the results of general relativity as the mass of the graviton goes to zero. However, a non-vanishing degree of freedom in the massless limit would lead to 
discontinuity found by van Dam and Veltman \cite{vanDam:1970vg} and
 Zakharov \cite{Zakharov:1970cc}. Although Vainshtein claimed that this problem can be cured by taking into account nonlinear interactions \cite{Vainshtein:1972sx}, Boulware and Daser pointed out the scalar degree of freedom responsible for Vainshtein's argument carries an extra ghost degree of freedom so-called BD ghost \cite{Boulware:1972aa}. Remarkably, this unwanted degree of freedom associated with higher derivatives can be eliminated by adding fully nonlinear graviton's mass terms such that those higher derivative terms vanish due to the total derivative terms, and this theoretically consistent massive 
gravity theory is now known as de Rham-Gabadadze-Tolley (dRGT) theory \cite{deRham:2010kj,deRham:2010ik}. 

In recent years, there have been a number of attempts for constructing a broad class of theories of massive gravity, e.g., mass-varying massive gravity \cite{Huang:2012pe}, quasi-dilation theory \cite{DAmico:2012hia},  and massive bi-gravity \cite{Hassan:2011zd}. The most interesting one would be new kinetic interactions for a massive graviton without introducing any extra degree of freedom. 
Such a kinetic interaction embedded in dRGT theory has been first found in the context of a pseudo-linear theory \cite{Hinterbichler:2013eza}, however, its nonlinear completions can not be included in dRGT theory in a consistent way due to the reappearance of BD ghost at non-linear level \cite{Kimura:2013ika,deRham:2013tfa}. The crucial problem of these derivative interactions was the appearance of higher derivatives of the scalar mode in Euler-Lagrange equations, and this, in general, leads to the Ostrogradsky's ghost \cite{Ostrogradsky1850}. 

Meanwhile it has been recently argued that higher derivatives in Euler-Lagrange equations are not essentially problematic as long as the appropriate number of constraints exists in a theory. Such concrete examples are found in the context of point particles~\cite{Langlois:2015cwa,Motohashi:2016ftl,Klein:2016aiq,Motohashi:2017eya,Kimura:2017gcy,Motohashi:2018pxg}, their field theoretical application~\cite{Crisostomi:2017aim,Kimura:2018sfs},  scalar-tensor theories~\cite{Gleyzes:2014dya,Langlois:2015cwa, 2015PhRvL.114u1101G,Crisostomi:2016czh, BenAchour:2016fzp}, vector-tensor theories~\cite{Kimura:2016rzw}. 
The key point of these theories is the degeneracy of the kinetic matrix, which provides an associated primary constraint (and subsequent constraint depending on the spin of a field), and an unwanted degree of freedom can be then successfully eliminated. In fact, in such a theory, higher time derivatives appearing in Euler-Lagrange equations should be removed by combining Euler-Lagrange equations and their time derivatives, 
therefore, initial conditions to solve the resultant differential equations are consistent with the number of degrees of freedom in ghost-free theories.

This fact opens up a new direction of study for searching new theories of a massive graviton, and it is therefore worth revisiting the pioneering attempt by Fierz and Pauli as a starting point of constructing a theoretically consistent massive gravity. To this end, in the present paper, we construct the most general quadratic theory of a massive spin-2 field and a massless spin-2 field with Lorentz invariance in flat spacetime, based on the Hamiltonian analysis. 
This paper is organized as follows. 
In Sec.~\ref{sec:theory}, we introduce the linear theory of a rank-2 symmetric tensor field
 and decompose it into tensors expressed by scalar quantities, those by vector quantities and those by tensor quantities, where each quantity is defined based on transformation properties with respect to a $3-$dimensional rotation in Minkowski spacetime.
The Hamiltonian formalism in Fourier space is also summarized. In Sec.~\ref{Sec:3}, we first derive the action for the transverse traceless tensor mode and a condition for avoiding instability. Then we find a condition to eliminate ghosty degrees of freedom for the scalar and vector modes and classify theories based on the Hamiltonian analysis. In Sec.~\ref{Sec:4}, we investigate the properties of obtained theories under the field redefinition to see relations with the known theories. 
Sec.~\ref{sec:summary} is devoted to summary.
In Appendix~\ref{sec:other cases}, we perform the complete Hamiltonian analysis of the special cases. In Appendix~\ref{BDproof}, the explicit proof of the existence of ghost degree of freedom is given if a theory has six or more degrees of freedom. In Appendix~\ref{App:gauge}, we derive gauge transformation and construct gauge invariant variables for each case. We also derive conditions for avoiding ghost and gradient instabilities for the scalar mode from reduced Lagrangian.

\section{Set up}
\label{sec:theory}

In this section, we introduce the most general Lorentz-invariant action for a rank-2 symmetric tensor field, which contains up to the Lagrangian quadratic in the tensor field and two derivatives with respect to spacetime.  
Since a theory for the rank-2 symmetric tensor field in general contains ten degrees of freedom, some of them might be ghost modes, which are unwanted degrees of freedom in a theory. 
To this end, we then apply scalar, vector, and tensor decomposition to the rank-2 symmetric tensor field which is defined based on transformation properties with respect to a $3-$dimensional rotation in Minkowski spacetime. We will also provide an overview of the Hamiltonian formalism in Fourier space.

\subsection{action}
Let us consider a generic Lorentz-invariant action for a rank-2 symmetric tensor field
$h_{\mu\nu}$ up to the quadratic order in Minkowski spacetime, 
\begin{align}
	S [h_{\mu \nu}] = \int {\rm d}^4 x \Bigl(-
	{\cal K}^{\alpha \beta | \mu \nu \rho \sigma} h_{\mu \nu, \alpha} h_{\rho \sigma, \beta}
	-{\cal M}^{\mu \nu \rho \sigma} h_{\mu \nu} h_{\rho \sigma} \Bigr) \,,
	\label{Lagrangian}
\end{align}
where ${\cal K}^{\alpha \beta | \mu \nu \rho \sigma}$ and ${\cal M}^{\mu \nu \rho \sigma}$ are  the most general combinations of the Minkowski metric 
$\eta_{\mu\nu}$, 
\begin{align}
	{\cal K}^{\alpha \beta | \mu \nu \rho \sigma}
	&= \kappa_1 \eta^{\alpha \beta} \eta^{\mu \rho} \eta^{\nu \sigma}
	+ \kappa_2 \eta^{\mu \alpha} \eta^{\rho \beta} \eta^{\nu \sigma}
	+ \kappa_3 \eta^{\alpha \mu} \eta^{\nu \beta} \eta^{\rho \sigma}
	+ \kappa_4 \eta^{\alpha \beta} \eta^{\mu \nu} \eta^{\rho \sigma} \,,\label{Ktensor}\\ 
	{\cal M}^{\mu \nu \rho \sigma}
	&= \mu_1 \eta^{\mu \rho} \eta^{\nu \sigma}
	+ \mu_2 \eta^{\mu \nu} \eta^{\rho \sigma} \label{Mtensor}\,,
\end{align}
and $\kappa_{1,2,3,4}$ and $\mu_{1,2}$ are constant parameters.
Contracting all the Minkowski metric, the action can be rewritten, after integration by parts, as
\begin{align}
	S [h_{\mu \nu}] = -\int {\rm d}^4 x \Bigl[ \kappa_1 h_{\mu \nu \,, \alpha} h^{\mu \nu \,, \alpha}
	+ \kappa_2 h^\alpha{}_{ \mu \,, \alpha} h^{\beta \mu}{}_{, \beta}
	+ \kappa_3 h^{\alpha \beta}{}_{, \alpha} h_{, \beta}
	+ \kappa_4 h_{, \alpha} h^{, \alpha}
	+ \mu_1 h_{\mu \nu} h^{\mu \nu} + \mu_2 h^2 \Bigr] \,,
\label{reLagrangian}
\end{align}
where the indices of $h_{\mu \nu}$ are raised by $\eta^{\mu \nu}$ and $h$ is the trace of $h_{\mu\nu}$ contracted with the Minkowski metric $\eta_{\mu\nu}$. 
A comma denotes a partial derivative with respect to spatial coordinates. 
The linearized Einstein-Hilbert action can be reproduced by setting
$\kappa_2 = -\kappa_3 = 2\kappa_4 = -2 \kappa_1$ up to overall factor,
and the kinetic term of \eqref{Lagrangian} is then invariant under the 
gauge transformation $h_{\mu\nu} \to h_{\mu\nu} + \partial_{\mu}\xi_{\nu}+ \partial_{\nu}\xi_{\mu}$, where $\xi_\mu$ is a gauge parameter. 
In addition to this choice of the parameters, when the mass parameters satisfy $ \mu_1 = -\mu_2 \neq 0$, 
the Lagrangian \eqref{Lagrangian} reproduces the Fierz-Pauli theory \cite{Fierz:1939ix}. Although the Fierz-Pauli theory respects the gauge invariance in the kinetic term, 
it is not necessary for a generic massive spin-2 field that we consider in the present paper.

\subsection{SVT decomposition}
In order to simplify analysis, 
we decompose the rank-2 symmetric tensor field $h_{\mu\nu}$ into 
a transverse-traceless tensor, tensors expressed by transverse vectors, and tensors expressed by scalars
where scalar, vector and tensors are defined based on transformation properties with respect to a $3-$dimensional rotation in Minkowski spacetime: 
\begin{align}
	h_{0 0} &= h^{0 0} = - 2 \alpha \,, \qquad
	h_{0 i} = - h^{0 i} = {\widehat{\beta}_{, i}} + B_i \quad (B^i{}_{, i} = 0) \\
	h_{i j} &= h^{i j}
	= 2 {\cal R} \delta_{i j} + 2\, {\widehat{{\cal E}}_{, i j}} + F_{i \,, j} + F_{j \,, i}
	+ 2 H_{i j} \quad ( F^i{}_{, i} = 0 \,, \quad H^i{}_i = H^{i j}{}_{, j} = 0) \,.
\end{align}
Here the transverse-traceless tensor $H_{ij}$, two transverse vectors $B_i$ and $F_i$, and four scalars $\alpha, \widehat\beta, {\cal R}$, and $\widehat{\cal E}$ respectively have two, four, and four components in total. Therefore, to obtain a theory whose
number of degrees of freedom is up to five including the degree of freedom of transverse-traceless tensor, we need to eliminate two components in the vector sector
and three components in the scalar sector respectively. Otherwise, ghost degree(s) of freedom appears as shown in Appendix~\ref{BDproof}.
Below we split the action into three parts and each of them is solely composed by single type of perturbations, that is, scalar, vector and tensor perturbations, which is always possible at the level of linear perturbation,
\begin{align}
	S [h_{\mu \nu}]
	&= S^S [\alpha \,, \widehat\beta \,, {\cal R} \,, \widehat{\cal E}\,] + S^V[B_i \,, F_i] + S^T [H_{i j}] \,.
\end{align}
In the next section, based on this separated action, we will look for theories with at most $5$ degrees of freedom finding the appropriate degeneracy conditions for the scalar and vector sectors, respectively.

\subsection{Hamiltonian formalism in Fourier space}
In this subsection, we briefly summarize the Hamiltonian formalism in the Fourier space. 
We, for convenience, work in the Fourier space, and the Fourier component of a field,  $A (t, \mathbf{k})$, is given by
\begin{align}
	A (t, \mathbf{k}) =
	\int {\rm d}^3 x A(t,\mathbf{x}) \mathrm{e}^{i \, \delta_{j k} k^j x^k}\,,
\end{align}
The Hamiltonian is defined by
\begin{eqnarray}
H (t) = \int {\rm d}^3 k \, {\cal H} (t, {\bf k}) \,,
\end{eqnarray}
where ${\cal H}$ is the Hamiltonian density in the Fourier space, 
\begin{eqnarray}
{\cal H} (t, {\bf k}) 
= \sum_I\dot s^I (t, {\bf k})  \, \pi_{s^I} (t,{\bf k}) - {\cal L} [s^I(t, {\bf k}), \dot{s}^I(t, {\bf k})] \,,
\end{eqnarray}
where $s^I$ and $\pi_{s^I}$ are respectively sets of canonical fields and their conjugate momenta. 
If the system has $n$ primary constraints ${\cal C}_i$, the total Hamiltonian and its density are given by 
\begin{eqnarray}
 H_T (t) = \int {\rm d}^3 k \, {\cal H}_T (t, {\bf k}) \,, \qquad 
{\cal H}_T (t, {\bf k})  = {\cal H} (t, {\bf k})  + \sum_{a=1}^n \lambda_{a} (t, {\bf k}) \, {\cal C}_a (t, {\bf k})\,,
\end{eqnarray}
where $\lambda_i$ are Lagrange multipliers associated with each primary constraint ${\cal C}_i$.
The Poisson bracket between ${\cal A}$ and ${\cal B}$ is defined by 
\begin{eqnarray}
\{{\cal A} (t, {\bf k}), {\cal B} (t, {\bf k}')\} = \int d^3k''
\sum_{I}
\left[
{\delta {\cal A} (t, {\bf k}) \over \delta s^I(t, {\bf k}'')} {\delta {\cal B} (t, {\bf k}') \over \delta \pi_{s^I}(t, {\bf k}'')}
-{\delta {\cal A} (t, {\bf k}) \over \delta \pi_{s^I}(t, {\bf k}'')} {\delta {\cal B} (t, {\bf k}') \over \delta s^I(t, {\bf k}'')}
\right] \,.
\end{eqnarray}
The time-evolution of the function ${\cal A}(t, {\bf k})$ is given by
\begin{eqnarray}
\dot{{\cal A}} (t, {\bf k}) = \{{\cal A} (t, {\bf k}), H_T (t)\} 
=  \int {\rm d}^3 k' \left[\{{\cal A} (t, {\bf k}), {\cal H} (t, {\bf k}')\} 
+ \sum_{a=1}^n \lambda_{a} (t, {\bf k}') \{{\cal A} (t, {\bf k}),  {\cal C}_a (t, {\bf k}') \} \right] \,.
\end{eqnarray}

\section{Hamiltonian analysis and classification}
\label{Sec:3}

In this section, we perform the Hamiltonian analysis for the theory (\ref{Lagrangian}) and classify the theory based on the number of degrees of freedom and constraint structures. 
Since the number of degrees of freedom in the theory is ten in general, one needs to adequately eliminate extra degrees of freedom in
each mode decomposed in the previous section. We first take a look at the tensor mode and derive the condition for avoiding a ghost mode. 
Then we seek conditions to eliminate unwanted modes for vector and scalar modes and conditions to have subsequent constraints. 
The existence of ghost degrees of freedom in the scalar sector is proved in the Appendix~\ref{BDproof} when the total number of degrees of freedom is more than five.

\subsection{Tensor modes}

The action in the tensor sector is given by
\begin{align}
	S^T [H_{i j}]
	&= 4 \int {\rm d} t \,{\rm d}^3 k \Bigl[  \kappa_1 \dot{H}_{ij}^2 - (\kappa_1 k^2 + \mu_1) H_{i j}^2 \Bigr] \,,
	\label{ST}
\end{align}
where a dot represents derivative with respect to time $t$.
As one can see from \eqref{ST}, the tensor modes are controlled by only two parameters $\kappa_1$ and $\mu_1$, and the existence of tensor modes and the condition for avoiding the ghost instability demand
\begin{align}
	``\text{Condition 1}"\ :\ \kappa_1 
	>0\,.
	\label{condition:1}
\end{align}
Throughout this paper, we always assume the "Condition 1" \eqref{condition:1}, and then the number of the degrees of freedom in the tensor sector is two. Furthermore, the parameter $\mu_1$ should not be negative
in order for avoiding the tachyonic instability in the tensor sector.

\subsection{Vector modes}
\label{sec:vector} 
In this subsection, we focus on the vector modes and find conditions to avoid extra ghost degrees of freedom, based on the Hamiltonian analysis. 
Before proceeding analysis, let us comment on counting the number of physical degrees of freedom and constraints in vector modes.
Since a vector in vector modes, $V_i$, satisfy the transverse condition, $k^i V_i = 0$, degrees of freedom in $V_i$ are two while there are three components. 
 Hence when we have primary constraints of vector type like $\pi_{V_i} = 0$, this should be understood as 
 two primary constraints and not three since the transverse conditions put a constraint for one component in $V_i$ and primary constraints should be obtained for the remaining two degrees of freedom.

The action for the vector modes 
can be written with the replacement $kF_i\to F_i$,
\begin{align}
	S^V [B_i \,, F_i]
	&= \int {\rm d} t \,{\rm d}^3 k \Bigl[ 
	-(2  \kappa_1+\kappa_2) \dot{B}_i^2 + 2 \kappa_1 \dot{F}_i^2 
      + 2 \kappa_2 k B_i \dot{F}_i
	+ 2 \left(\kappa_1 k^2+\mu_1\right) B_i^2 -\left(k^2 (2 \kappa_1+\kappa_2) 
	+ 2 \mu_1\right) F_i^2
	\Bigr] \,.
	\label{SV}
\end{align}
One may immediately notice from (\ref{SV}) that there are the appearance of either ghost or gradient instabilities in $B_i$ or $F_i$ modes, depending on the sign of $2\kappa_1+\kappa_2$ as well as $\kappa_1$. This concludes that one needs to at least eliminate either $B_i$ or $F_i$ in order to have two degrees of freedom. The existence of the tensor modes \eqref{condition:1} leads to the unique option to have a primary constraint for $B_i$, that is, 
\begin{align}
	``\text{Condition 2}"\ :\ 
	2 \kappa_1 + \kappa_2 = 0 \qquad
	\Longleftrightarrow  \qquad
	\kappa_2 = -2\kappa_1 \,.
	\label{condition:2}
\end{align}
With this condition \eqref{condition:2}, the kinetic term of $B_i$ vanishes, which implies $B_i$ are manifestly non-dynamical. 
Then the action for the vector mode can be recast as 
\begin{align}
	S^V [B_i \,, F_i]
	&= \int{\rm d} t\, {\rm d}^3 k\,{\cal L}^V
	= \int {\rm d} t\, {\rm d}^3 k \Bigl[ 
	 2 \kappa_1 \dot{F}_i^2 - 4 \kappa_1 k B_i \dot{F}_i 
	+ 2 \left(\kappa_1 k^2
	+ \mu_1\right) B_i^2 - 2 \mu_1 F_i^2
	\Bigr] \,.
\end{align}
Apparently, the action for the vector modes depends on only two parameters, 
$\kappa_1$ and $\mu_1$, as in the tensor modes. 
The conjugate momenta for $B_i$ and $F_i$ are given by
\begin{align}
\pi_{B_i} &\equiv \frac{\delta {\cal L}^V}{\delta \dot{B}_i} = 0 \,,\\
\pi_{F_i} &\equiv \frac{\delta {\cal L}^V}{\delta \dot{F}_i} =  4 \kappa_1 (\dot{F}_i - k B_i) \,,
\end{align}
and we therefore have two primary constraints instead of three as mentioned at the beginning of this subsection, which are defined by
\begin{align}
	{\cal C}_1^{B_i} = \pi_{B_i} = 0 \,.
	\label{primaryBi}
\end{align}
Then Hamiltonian and the total Hamiltonian densities read
\begin{align}
	{\cal H}^V &= \dot{B}_i \pi_{B_i} + \dot{F}_i \pi_{F_i} - {\cal L}^V 
	\approx \frac{\pi_{F_i}^2}{8 \kappa_1} + k \pi_{F_i} B_i - 2 \mu_1 B_i^2 + 2 \mu_1 F_i^2\,,\nonumber\\
	{\cal H}_T^V & = {\cal H}^V + \lambda_{B_i} \pi_{B_i} \,,
\end{align}
where $\lambda_{B_i}$ are Lagrange multipliers.
We have suppressed the terms $\dot{B}_i \pi_{B_i}$ in the final expression of ${\cal H}^V$ since they vanish once the primary constraints (\ref{primaryBi}) are imposed.
One can easily check that the evolution of the primary constraints automatically yields secondary constraints,
\begin{align}
	{\cal C}_2^{B_i} \equiv \dot{{\cal C}}_1^{B_i} = \{ {\cal C}_1^{B_i} \,, H_T^V \}
	 = \{ {\cal C}_1^{B_i} \,, H^V \}
	=  k \pi_{F_i} + 4 \mu_1 B_i \approx 0\,.
\end{align}
Then the time-evolution of the secondary constraints are given by\footnote{
To be precise, one needs the integral over the Fourier space 
in front of the Lagrange multipliers $\lambda_{B_j}$, which can be always 
integrable because of the appearance of the Dirac's delta function. For simplicity, we omit this integral and arguments of each variable since
results do not change.
}
\begin{align}
	\dot{{\cal C}}_2^{B_i} = \{ {\cal C}_2^{B_i} \,, H_T^V \} = \{ {\cal C}_2^{B_i} \,, H^V \} 
	+ \{ {\cal C}_2^{B_i} \,, {\cal C}_1^{B_j} \} \lambda_{B_j} \approx 0\,,
	\label{dotC2B}
\end{align}
where the coefficients of $\lambda_{B_i}$ are given by
\begin{align}
	\{ {\cal C}_2^{B_i} \,, {\cal C}_1^{B_j} \} &= 4\mu_1 \,\delta_{ij} \,.
\label{C2BC1B}
\end{align}
Therefore, we have two cases :

\vspace{2mm}

{\bf Case V1 :} $\mu_1 = 0$ \\
In this case, in addition that the Poisson bracket \eqref{C2BC1B} vanishes, $\{ {\cal C}_2^{B_i}\,, H^V\}$ is trivially zero. Then, there is no more constraint.  
Thus there are two primary constraints ${\cal C}_1^{B_i}$ and two secondary constraints ${\cal C}_2^{B_i}$, and 
all of them are first-class since all the Poisson brackets between these constraints vanish. Therefore,  
\begin{eqnarray}
\textrm{vector DOF} = \frac{4\times 2 - 4 \,(\textrm{2 primary \& 2 secondary}) \times 2 \,(\textrm{first-class})}{2} 
=0 \,.
\end{eqnarray}
This case is exactly the same as the linearized Einstein's gravity, and thus the Lagrangian is invariant under the gauge transformation, $B_i \to B_i + {\dot \zeta}_i$ and $F_i \to F_i + \zeta_i$, where $\zeta_i$ is 3-vector satisfying the transverse condition $\partial^i \zeta_i =0$.
Note that a theory with only one degree of freedom in the vector sector is prohibited by spatial covariance of the theory.

\vspace{2mm}

{\bf Case V2 :} $\mu_1 \neq 0$ \\
When $\mu_1 \neq 0$, the last equation \eqref{dotC2B} can be used to determine the Lagrange multipliers $\lambda_{B_i}$, 
\begin{align}
\lambda_{B_i} \approx 
{1 \over 4\mu_1}\{ {\cal C}_2^{B_i} \,, H^V \} \,.
\end{align}
Thus, there are two primary constraints ${\cal C}_1^{B_i}$ and two secondary constraints ${\cal C}_2^{B_i}$, and 
all of them are second-class since the Poisson brackets between these constraints are non-vanishing. 
Therefore, the number of the degrees of freedom for the vector modes is given by 
\begin{eqnarray}
\textrm{vector DOFs} = \frac{4\times 2 - 4 \,(\textrm{2 primary \& 2 secondary})}{2}
=2\,.
\end{eqnarray}
This case includes the FP theory.

\subsection{Scalar modes}
\label{sec:scalar}
In this subsection, we investigate the scalar modes and classify theories by finding condition to avoid appearing extra ghost degrees of freedom. In the Appendix~\ref{BDproof}, we see that there are dangerous degrees of freedom if the system has two or more degrees of freedom. We also derive gauge transformation and conditions for avoiding instabilities of obtained theories in the Appendix~\ref{App:gauge}.

As for scalar perturbations, by introducing dimensionless quantities $\beta$ and ${\cal E}$
 which are defined by $\beta \equiv k \widehat{\beta}$ and ${\cal E} \equiv k^2 \widehat{{\cal E}}$, respectively, 
 the action reduces to
\begin{align}
	S^S [\alpha \,, \beta \,, {\cal R} \,, {\cal E}]
	&=\int{\rm d} t\, {\rm d}^3 k\,{\cal L}^S
	= \int {\rm d} t \, {\rm d}^3 k\, \Bigl( {\cal L}_\mathrm{kin}^S + {\cal L}_\mathrm{cross}^S + {\cal L}_\mathrm{mass}^S \Bigr) \,,
	\label{SS}
\end{align}
where 
\begin{align}
	{\cal L}_\mathrm{kin}^S &=
	 4 (\kappa_1+\kappa_2+\kappa_3+\kappa_4) \dot{\alpha}^2
	- (2 \kappa_1+\kappa_2) \dot\beta^2 
	+ 12 (\kappa_1+3 \kappa_4) \dot{{\cal R}}^2
	+ 4 (\kappa_1+\kappa_4) \dot{{\cal E}}^2 
	\notag\\
	& \qquad
	- 4 (\kappa_3+2 \kappa_4)\left( -3\dot{{\cal R}} + \dot{{\cal E}} 
	\right) \dot{\alpha}
	-8 (\kappa_1+3 \kappa_4) \dot{{\cal R}} \dot{{\cal E}} 
	\,,
	\label{Skin}\\
	{\cal L}_\mathrm{cross}^S &=
	-4
	\left[  (\kappa_2+\kappa_3) \dot{\alpha}
	+  (\kappa_2+3 \kappa_3) \dot{{\cal R}}
	-  (\kappa_2+\kappa_3) \dot{\cal E} 
	\right] k \beta \,,
	\\
	{\cal L}_\mathrm{mass}^S &=
	-4 \Bigl[ k^2 (\kappa_1+\kappa_4)+\mu_1+\mu_2\Bigr] \alpha^2
	+ \Bigl[  k^2 (2 \kappa_1+\kappa_2) +2 \mu_1 \Bigr] \beta^2 
	\notag\\
	& \qquad
	- 4\Bigl[ k^2 (3 \kappa_1+\kappa_2+3 \kappa_3+9 \kappa_4)+3 (\mu_1+3 \mu_2) \Bigr] {\cal R}^2
	- 4 \Bigl[ k^2 (\kappa_1+\kappa_2+\kappa_3+\kappa_4)+\mu_1+\mu_2 \Bigr] {\cal E}^2 
	\notag\\
	& \qquad
	- 4\Bigl[ \Bigl( k^2 (\kappa_3+6 \kappa_4)+6 \mu_2 \Bigr) {\cal R}
	- \Bigl( k^2 (\kappa_3+2 \kappa_4)+2 \mu_2 \Bigr) {\cal E} 
	\Bigr] \alpha
	\notag\\
	& \qquad
	+8 \Bigl[ k^2 (\kappa_1+\kappa_2+2 \kappa_3+3 \kappa_4)+ (\mu_1+3 \mu_2) \Bigr] {\cal R} {\cal E} 
	\,.\label{LS mass term}
\end{align}
Now under the ``Condition 2'' \eqref{condition:2}, $2 \kappa_1 + \kappa_2 = 0$, the time derivative of $\beta$ vanishes in the Lagrangian. Then $\beta$ becomes non-dynamical. 

The canonical momenta for $Q = \{ \alpha\,, \beta \,, {\cal R} \,, {\cal E} \}$ are defined by $\pi_Q \equiv \delta {\cal L}^S/\delta \dot{Q}$ and read
\begin{align}
\begin{pmatrix}
\pi_\alpha \\
\pi_\beta \\ 
\pi_{\cal R} \\  
\pi_{\cal E} \\
\end{pmatrix}
 = 4 \, {\cal K}^S \,
\begin{pmatrix}
 \dot{\alpha} \\
 \dot{\beta} \\ 
 \dot{{\cal R}} \\  
 \dot{{\cal E}} \\
\end{pmatrix}
 + 4
\begin{pmatrix}
2\kappa_1 - \kappa_3 \\
0 \\
2\kappa_1-3\kappa_3 \\
- 2\kappa_1+\kappa_3
\end{pmatrix}
k\beta \,,
\end{align}
 where
\begin{align}
 {\cal K}^S \equiv
\begin{pmatrix}
2(- \kappa_1 +\kappa_3 +\kappa_4) & 0 &
3(\kappa_3+2\kappa_4) &
- (\kappa_3+2\kappa_4) \\
0 & 0 & 0 & 0 \\
3 (\kappa_3+2\kappa_4) & 0 &
6 (\kappa_1+3\kappa_4) &
- 2 (\kappa_1+3\kappa_4) \\
- (\kappa_3+2\kappa_4) & 0 &
- 2 (\kappa_1+3\kappa_4) &
2 (\kappa_1+\kappa_4) 
\end{pmatrix}
 \,.
\label{KineticMatrix33}
\end{align}
Let us calculate the determinant of the kinetic matrix, ${\cal K}^S$, which is given by
\begin{align}
	|{\cal K}^S| = 
	- 4 \kappa_1 \left(4 \kappa_1^2-4 \kappa_1 \kappa_3+8 \kappa_1
	\kappa_4+3 \kappa_3^2\right) \,.
\end{align}
Taking into account the condition for having the tensor mode, $\kappa_1\neq0$,
the determinant vanishes only when 
\begin{align}
	``\text{Condition 3}" \ :\ 
	4 \kappa_1^2-4 \kappa_1 \kappa_3+8 \kappa_1
	\kappa_4+3 \kappa_3^2 =0
	\qquad \Longleftrightarrow \qquad
	\kappa_4 = - \frac{4 \kappa_1^2 - 4 \kappa_1 \kappa_3 + 3 \kappa_3^2}{8 \kappa_1} \,.
	\label{condition:3}
\end{align}
Then, the degeneracy of the kinetic matrix \eqref{KineticMatrix33} leads to an additional primary constraint in addition to the one for $\beta$, $\pi_\beta = 0$. 
Note that the linearized Einshtein-Hilbert kinetic term satisfies the ``Condition 3" \eqref{condition:3}. 
When this degeneracy condition is satisfied, one of the eigenvalues of the kinetic matrix vanishes
 and the remaining eigenvalues, $\lambda$, will be a solution of the following eigen equation:
\begin{eqnarray}
 \lambda^2 + \left( 10 \kappa_1 - 26 \kappa_3 + \frac{33 \kappa_3^2}{2 \kappa_1} \right) \lambda
 = 8 \Bigl[ 4 (\kappa_1 - \kappa_3)^2 + \kappa_3^2 \Bigr] > 0\,.
\end{eqnarray}
As long as $\kappa_{1,3}$ are real, other eigenvalues will be non-vanishing.

\subsubsection{
one primary constraint :  $4 \kappa_1^2-4 \kappa_1 \kappa_3+8 \kappa_1\kappa_4+3 \kappa_3^2 \neq 0$}
\label{sec:VII}

Let us consider the case with only one primary constraint, i.e., the ``Condition 3'' (\ref{condition:3}) is not imposed.
Thus, we have a primary constraint :
\begin{align}
{\cal C}^\beta_1 = \pi_\beta = 0 \,.
\label{primaryconstraint1}
\end{align}
The Hamiltonian and the total Hamiltonian densities in the scalar sector reads
\begin{align}
{\cal H}^S &= \dot{\alpha} \pi_\alpha + \dot{\beta} \pi_\beta + \dot{{\cal R}} \pi_{\cal R} + \dot{{\cal E}} \pi_{\cal E} - {\cal L}^S  \nonumber \\
&\approx
4\left[\mu_1+\mu_2 + k^2(\kappa_1+ \kappa_4)\right]\alpha^2 -2\mu_1 \beta^2 
+4\left[\mu_1 +\mu_2 -k^2(\kappa_1-\kappa_3-\kappa_4)\right] {\cal E}^2\nonumber \\
&~~~
-8\left[\mu_1+3\mu_2 -k^2(\kappa_1-2\kappa_3-3\kappa_4)\right]{\cal E}{\cal R}
+4\left[3\mu_1 +9\mu_2 +k^2(\kappa_1+3\kappa_3 +9\kappa_4)\right]{\cal R}^2\nonumber \\
&~~~
-\left[4\left(2\mu_2 + k^2 (\kappa_3+2\kappa_4)\right){\cal E} -4 \left(6\mu_2+k^2(\kappa_3+6\kappa_4)\right){\cal R}\right]\alpha\nonumber \\
&~~~
+{1 \over 32 \kappa_1} \left(2\pi_{\cal R} + 3\pi_{\cal E}  \right) \pi_{\cal E}+(\pi_\alpha + \pi_{\cal E})k\beta
-\frac{8{\kappa_1}(\kappa_1+3\kappa_4) \pi_\alpha^2 -8{\kappa_1}(\kappa_3+2\kappa_4)\pi_\alpha \pi_{\cal R}
	- (2\kappa_1 - \kappa_3)^2 \pi_{\cal R}{^2}
}{32{\kappa_1}(4\kappa_1^2-4\kappa_1\kappa_3+3\kappa_3^2+8\kappa_1\kappa_4)}
\,,  \\
{\cal H}_T^S&
= {\cal H}^S + \lambda_\beta \pi_\beta \,.
\end{align}
We have suppressed the term $\dot{\beta} \pi_\beta$ in the final expression of ${\cal H}^S$ since it vanishes once the primary constraint (\ref{primaryconstraint1}) is imposed.
The evolution of the primary constraint automatically yields a secondary constraint
\begin{align}
{\cal C}^\beta_2 \equiv \dot{{\cal C}}^\beta_1 = \{ {\cal C}^\beta_1 \,, H_T^S \} = \{ {\cal C}^\beta_1 \,, H^S \}
=  -k\pi_\alpha -  k\pi_{\cal E} +4 \mu_1 \beta \approx 0\,.
\label{secondaryconstraint}
\end{align}
Then, the Poisson bracket between the secondary and primary constraints of $\beta$ is given by
\begin{align}
\{ {\cal C}^\beta_2 \,, {\cal C}^\beta_1 \} &= 4\mu_1 \,.
\end{align}
If $\mu_1 \neq 0$, then no more constraint will be generated and the last equation can be used to determine the value of $\lambda_\beta$, that is 
$\lambda_\beta \approx - \{ {\cal C}^\beta_2 \,, H^S \} / \{ {\cal C}^\beta_2 \,, {\cal C}^\beta_1 \} $.
In this case, we have one primary constraint ${\cal C}_1^\beta$, and one secondary constraint ${\cal C}_2^\beta$, all of which are 
second-class. Therefore, the number of degrees of freedom in the scalar sector is $(8-2)/2=3$,
signalling the existence of an extra degree of freedom. The explicit proof of the existence of a ghost degree of freedom is 
given in the Appendix \ref{BDproof}.
Therefore, one has to impose an extra condition $\mu_1 = 0$ in order to 
eliminate this extra degree of freedom.

\vspace{2mm}
{\bf Case SI :} $\mu_1 = 0$, $4\kappa_1^2-4\kappa_1\kappa_3+3\kappa_3^2+8\kappa_1\kappa_4 \neq 0$ \\
If $\mu_1=0$, the consistency of the secondary constraint ${\cal C}_2^\beta$ generates a tertiary constraint, 
\begin{eqnarray}
{\cal C}_3^\beta \equiv \dot{{\cal C}}_2^\beta 
= \{ {\cal C}^\beta_2 \,, H_T^S \} = \{ {\cal C}^\beta_2 \,, H^S \} 
= 4k^3 \bigl[
(2\kappa_1 - \kappa_3) \alpha +(2\kappa_1 -3 \kappa_3) {\cal R} -(2\kappa_1-\kappa_3){\cal E}
\bigr] \approx 0\,.
\end{eqnarray}
Now one can check that all the constraints commute each other, i.e., $\{ {\cal C}^\beta_i \,, {\cal C}^\beta_j \}=0$ where ($i,j = 1,2,3$). One can also check that $\dot{{\cal C}}_3^\beta=k^2 {\cal C}_2^\beta$, implying no more constraint.
There are one primary constraint ${\cal C}_1^\beta$, one secondary constraint ${\cal C}_2^\beta$, and one tertiary constraint ${\cal C}_3^\beta$, and all of them 
are first-class. Therefore, 
\begin{eqnarray}
\textrm{scalar DOF} = \frac{4\times 2 - 3\,(\textrm{1 primary \& 1 secondary \& 1 tertiary})\times 2 \,(\textrm{first class})}{2}
=1
\end{eqnarray}
Since $\mu_1=0$ corresponds to Case V1, the vector sector does not have any degree of freedom, and the total number of degrees of freedom in this case is three.

\subsubsection{two primary constraints : $4 \kappa_1^2-4 \kappa_1 \kappa_3+8 \kappa_1
	\kappa_4+3 \kappa_3^2=0$} 
Hereafter we impose the ``Condition 3'' (\ref{condition:3}), implying the existence of two primary constraints.
As one can easily see, the canonical momenta $\pi_\alpha$ and $\pi_{{\cal R}}$ are 
not independent because of the degeneracy condition \eqref{condition:3}, 
i.e., 
\begin{eqnarray}
 (2\kappa_1 - 3\kappa_3) \pi_\alpha - (2\kappa_1-\kappa_3) \pi_{{\cal R}} = 0 \,.
\label{primaryconstraint2}
\end{eqnarray}
This clearly shows that there are two primary constraints in this class of theories, namely \eqref{primaryconstraint1} and \eqref{primaryconstraint2}. 
We classify theories depending on vanishing or non-vanishing of the coefficient of $\pi_\alpha$, 
$2\kappa_1 - 3\kappa_3$. 
Hereafter, we will focus only on the case of $2\kappa_1-3\kappa_3\neq 0$. 
The Hamiltonian analysis in the class of models with $2\kappa_1 - 3\kappa_3=0$ is presented
 in Appendix \ref{sec:other cases}.

In this case, we have the following two primary constraints :
\begin{align}
	{\cal C}^\alpha_1 &= \pi_\alpha -{2\kappa_1-\kappa_3 \over 2\kappa_1-3\kappa_3} \pi_{\cal R} = 0 \,, \qquad
	{\cal C}^\beta_1 = \pi_\beta = 0 \,.
	\label{primaryalphabeta}
\end{align}
Here it is clear that ${\cal C}_1^\alpha$ and ${\cal C}_1^\beta$ commute each other.
Then the Hamiltonian and the total Hamiltonian densities read
\begin{align}
	{\cal H}^S &=\dot{\alpha} \pi_\alpha + \dot{\beta} \pi_\beta + \dot{{\cal R}} \pi_{\cal R} + \dot{{\cal E}} \pi_{\cal E} - {\cal L}^S \,,\nonumber \\
	&\approx \left[\frac{k^2 (2\kappa_1-\kappa_3)(2\kappa_1+3\kappa_3)}{2\kappa_1}+4(\mu_1+\mu_2) \right] \alpha^2  
	- 2\mu_1 \beta^2 
	+ \left[- \frac{3k^2(2\kappa_1-\kappa_3)^2}{2\kappa_1}+4(\mu_1+\mu_2)\right]{\cal E}^2\nonumber\\
	&~~~
	+ \left[\frac{k^2(2\kappa_1-\kappa_3)(10\kappa_1-9\kappa_3)}{\kappa_1}-8(\mu_1+3\mu_2)\right]{\cal E}{\cal R}
	+\left[- \frac{k^2(14\kappa_1-9\kappa_3)(2\kappa_1-3\kappa_3)}{2\kappa_1}+12(\mu_1+3\mu_2)\right]{\cal R}^2 \nonumber\\
	&~~~
	+ \left[\frac{k^2(2\kappa_1-\kappa_3)(2\kappa_1-3\kappa_3)}{\kappa_1}-8\mu_2\right]\alpha {\cal E} 
	+ \left[ - \frac{k^2 (12\kappa_1^2 -16\kappa_1\kappa_3 + 9\kappa_3^2)}{\kappa_1} + 24\mu_2\right]\alpha{\cal R}  \nonumber\\
	&~~~ 
	+k \beta\pi_{\cal E}+\frac{3}{32 \kappa_1}\pi_{\cal E}^2 
	+\frac{2\kappa_1-\kappa_3}{2\kappa_1-3\kappa_3}k\beta\pi_{\cal R} 
	+{1 \over 16 \kappa_1}\pi_{\cal R} \pi_{\cal E}
	-\frac{(2\kappa_1-\kappa_3)(2\kappa_1+3\kappa_3)}{32 \kappa_1 (2\kappa_1 -3\kappa_3)^2}\pi_{\cal R}^2 
	\,,
\end{align}
 and
\begin{align}
	{\cal H}_T^S&= {\cal H}^S + \lambda_\alpha {\cal C}_1^\alpha + \lambda_\beta {\cal C}_1^\beta \,.
\end{align}
where we have suppressed the terms proportional to the primary constraints (\ref{primaryalphabeta}) in the final expression of ${\cal H}^S$.

The evolution of the primary constraints requires
\begin{align}
	{\cal C}_2^\alpha &\equiv \dot{{\cal C}}_1^\alpha 
	= \{ {\cal C}^\alpha_1 \,, H_T^S \} = \{ {\cal C}^\alpha_1 \,, H^S \} = c_{2}^\alpha\, \alpha + c_{2}^{\cal R}\, {\cal R} + c_{2}^{\cal E}\, {\cal E} \approx 0\,,
\label{const:c2alpha}	
	\\
	{\cal C}_2^\beta &\equiv\dot{{\cal C}}_1^\beta 
	= \{ {\cal C}^\beta_1 \,, H_T^S \} = \{ {\cal C}^\beta_1 \,, H^S \} = -\frac{2\kappa_1 - \kappa_3}{2\kappa_1-3\kappa_3}k \pi_{\cal R} 
	- k\pi_{\cal E} +4\mu_1 \beta \approx 0\,,
\label{const:c2beta}	
\end{align}
where we have defined the coefficients, 
\begin{eqnarray}
c_2^\alpha&=&-\frac{8\left[\mu_1 (2\kappa_1 -3\kappa_3) -4\mu_2 \kappa_1 + 2k^2(2\kappa_1-\kappa_3)(\kappa_1-\kappa_3)\right]}{2\kappa_1 -3\kappa_3} \,,
\label{coe:c2a}
\\
c_2^{\cal R}&=&-\frac{8\left[-3\mu_1 (2\kappa_1 -\kappa_3) -12\mu_2 \kappa_1 + 2k^2(2\kappa_1-3\kappa_3)(\kappa_1-\kappa_3)\right]}{2\kappa_1 -3\kappa_3} \,,
\label{coe:c2r}
\\
c_2^{\cal E}&=&-\frac{8\left[\mu_1 (2\kappa_1 -\kappa_3) +4\mu_2 \kappa_1 - 2k^2(2\kappa_1-\kappa_3)(\kappa_1-\kappa_3)\right]}{2\kappa_1 -3\kappa_3} \,.
\label{coe:c2e}
\end{eqnarray}
In general we will have two secondary constraints, but there exists an exceptional case where only one secondary constraint exists if $\kappa_1 = \kappa_3$. With this condition satisfied, the terms proportional to $k^2$ in the numerator of
 Eqs.~(\ref{coe:c2a}, \ref{coe:c2r}, \ref{coe:c2e}) vanish and hence ${\cal C}_2^\alpha$ reduces to a trivial equation depending on the value of $\mu_1$ and $\mu_2$. This special case happens either $\mu_1 = \mu_2 =0$ or $\mu_1+4\mu_2 = 0$ is satisfied, which will be investigated in the end of this subsection. 
 In the meantime, we consider the case with two non-trivial secondary constraints assuming $\kappa_1 \neq \kappa_3$.

The Poisson brackets between the primary and secondary constraints are given by
\begin{align}
	\{ {\cal C}_2^\alpha \,, {\cal C}_1^\alpha \} &=- \frac{32}{(2\kappa_1-3\kappa_3)^2}\left[\mu_1(4\kappa_1^2 -6\kappa_1 \kappa_3 + 3\kappa_3^2)+4\mu_2\kappa_1^2\right]\,,\label{C2aC1a} \\
	\{ {\cal C}_2^\beta \,, {\cal C}_1^\beta \} &= 4 \mu_1 \,. \label{C2bC1b}
\end{align}
Thus, as long as these Poisson brackets are non-vanishing, 
no more constraints will be generated.
In this case, we have two primary constraints ${\cal C}_1^\alpha, {\cal C}_1^\beta$ and two secondary constraints ${\cal C}_2^\alpha, {\cal C}_2^\beta$ , and 
all the constraints are second-class, which implies the number of degrees of freedom 
is $(8-4)/2=2$. The explicit proof of the existence of a ghost degree of freedom is 
given in the Appendix \ref{BDproof}. 
Therefore, one needs to eliminate an extra degree of freedom. 
In order to have an extra constraint, there are three options : 
both $\mu_1$ and $\mu_2$ vanish ({\bf Case SIIa}), either \eqref{C2aC1a} ({\bf Case SIIc}) 
or \eqref{C2bC1b} ({\bf Case SIIb}) vanishes.

\vspace{2mm}

{\bf Case SIIa :} $\mu_1=\mu_2=0$, 
\\
If $\mu_1=\mu_2=0$, both \eqref{C2aC1a} and \eqref{C2bC1b} vanish. 
In addition,one can see
$\dot{\cal C}_2^\alpha \propto {\cal C}_2^\beta \approx 0$ and
$\dot{\cal C}_2^\beta \propto {\cal C}_2^\alpha \approx 0$\,,
which implies that no more constraint is therefore generated.
Since all the constraints commute, all the primary and secondary constraints are first-class. 
Therefore,
\begin{eqnarray}
\textrm{scalar DOF} = \frac{4\times 2 - 4 \,(\textrm{2 primary \& 2 secondary})\times 2 \,(\textrm{first-class})}{2}
=0 \,.
\end{eqnarray}
In this case,
the scalar mode as well as the vector mode do not have any DOF (Case V1), and only the tensor mode
can propagate. 
Since the linearized Einstein-Hilbert term satisfy the ``Condition 3'' \eqref{condition:3}, and the mass terms are absent, this class reduces to linearized general relativity when $2\kappa_1-\kappa_3=0$. 
As we will see in the next section, the whole parameter family of this case SIIa can be mapped from linearized general relativity by a field redefinition.

\vspace{2mm}

{\bf Case SIIb :} $\mu_1 = 0$, $\mu_2 \neq 0$, \\
In this case, since \eqref{C2aC1a} is non-vanishing while \eqref{C2bC1b} vanishes, 
$\lambda_\alpha$ is determined by $\dot{{\cal C}}^\alpha_2 \approx 0$, 
that is  $\lambda_\alpha = -\{{\cal C}_2^\alpha, H^S\}/\{{\cal C}_2^\alpha, {\cal C}_1^\alpha\}$. 
On the other hand as for ${\cal C}_2^\beta$, 
since ${\cal C}_2^\beta$ does not commute with ${\cal C}_2^\alpha$, 
we shall consider a linear combination of ${\cal C}_2^\beta$ and ${\cal C}_1^\alpha$
which commute with ${\cal C}_2^\alpha$
instead of the original ${\cal C}_2^\beta$:
\begin{eqnarray}
 \widetilde{{\cal C}}_2^\beta &\equiv {\cal C}_2^\beta - k \, {\cal C}_1^\alpha = - k (\pi_\alpha + \pi_{\cal E})\,.
\end{eqnarray}
The consistency of $\widetilde{{\cal C}}^\beta_2$ yields a tertiary constraint,
\begin{eqnarray}
{\cal C}_3^\beta &\equiv & \dot{\widetilde{{\cal C}}}{}_2^\beta 
= \{ \widetilde{{\cal C}}^\beta_2 \,, H_T^S \} = \{ \widetilde{{\cal C}}^\beta_2 \,, H^S \}
= 
4 k^3 \Bigl[ (2 \kappa_1 - \kappa_3) \alpha + (2 \kappa_1 - 3 \kappa_3) {\cal R}
 - (2 \kappa_1 - \kappa_3) {\cal E} \Bigr]
\approx 0 \,.
\label{eq:C3b}
\end{eqnarray}
The Poisson brackets between this constraint and primary constraints vanish
and one also find $\dot{\cal C}_3^\beta = k^2{\cal C}_2^\beta = k^2 (\widetilde{{\cal C}}_2^\beta + k {\cal C}_1^\alpha) \approx 0$ 
implying that no more constraint is generated.
One can straightforwardly show that the constraints ${{\cal C}}^\beta_1$, ${\widetilde {\cal C}}_2^\beta$, and 
${{\cal C}}_3^\beta$ 
commute with all constraints including themselves. 
Therefore,  there are two primary constraints ${\cal C}_1^\alpha, {\cal C}_1^\beta$, two secondary constraints ${\cal C}_2^\alpha, {\widetilde {\cal C}}_2^\beta$, and one tertiary constraint ${\cal C}_3^\beta$. The constraints ${{\cal C}}^\beta_1$, ${\widetilde {\cal C}}_2^\beta$, and ${\cal C}_3^\beta$ are first-class, and rest of them are second-class. Therefore, 
\begin{eqnarray}
\textrm{scalar DOF} = \frac{4\times 2 - 3 \, (\textrm{1 primary \& 1 secondary \& 1 tertiary}) \times 2 \, (\textrm{first-class})- 2 \, (\textrm{1 primary \& 1 secondary})}{2}
=0 \,. \nonumber\\
\end{eqnarray}
In this case, the vector mode does not propagate (Case V1) and the total number of degrees of freedom is two.

\vspace{2mm}

{\bf Case SIIc :} $\mu_1(4\kappa_1^2 -6\kappa_1 \kappa_3 + 3\kappa_3^2)+4\mu_2\kappa_1^2 =0$, $\mu_1 \neq 0$, \\
In this case, since \eqref{C2bC1b} does not vanish, 
$\lambda_\beta$ can be determined by imposing $\dot{{\cal C}}^\beta_2\approx 0$,
that is  $\lambda_\beta = -\{{\cal C}_2^\beta, H^S\}/\{{\cal C}_2^\beta, {\cal C}_1^\beta\}$,
and hence no more constraint will be generated as for ${\cal C}_1^\beta$. 
On the other hand, we solve the condition for $\mu_2$ from $\{{\cal C}_2^\alpha, {\cal C}_1^\alpha\} = 0$, which is given by
\begin{eqnarray}
	``\text{Condition 4}" \ :\ \mu_2 
	= - \frac{4 \kappa_1^2-6 \kappa_1 \kappa_3+3 \kappa_3^2}{4 \kappa_1^2} \mu_1 \,.
\label{condition:4}
\end{eqnarray} 
Since ${\cal C}_2^\alpha$ does not commute with ${\cal C}_2^\beta$, 
 we shall consider a linear combination of ${\cal C}_2^\alpha$ and ${\cal C}_1^\beta$ 
which commutes with ${\cal C}_2^\beta$
instead of the original ${\cal C}_2^\alpha$:
\begin{align}
 \widetilde{{\cal C}}_2^\alpha 
 &\equiv {\cal C}_2^\alpha + k \frac{8 (\kappa_1 - \kappa_3)}{2 \kappa_1 - 3 \kappa_3} {\cal C}_1^\beta \,.
\end{align}
Then, the time consistency of $\widetilde{{\cal C}}^\alpha_2$ yields a tertiary constraint
\begin{align}
 {\cal C}_3^\alpha \equiv \dot{\widetilde{{\cal C}}}{}_2^\alpha 
 = \{ \widetilde{{\cal C}}^\alpha_2 \,, H_T^S \} = \{ \widetilde{{\cal C}}^\alpha_2 \,, H^S \} 
 =
 - \frac{4 (\kappa_1 - \kappa_3)}{2 \kappa_1 - 3 \kappa_3}
 \left( \frac{k^2 (2 \kappa_1 - \kappa_3) - 2 \mu_1}{2 \kappa_1 - 3 \kappa_3} \pi_{\cal R} 
 + k^2 \pi_{\cal E} \right)
  \approx 0 \,.
\label{eq:C3a}
\end{align}
The Poisson bracket between this constraint and primary constraints vanishes and the 
 time-consistency of ${\cal C}_3^\alpha$ yields a quaternary constraint
\begin{eqnarray}
{\cal C}_4^\alpha \equiv \dot{{\cal C}}_3^\alpha 
= \{ {\cal C}^\alpha_3 \,, H_T^S \} = 
\{ {\cal C}^\alpha_3 \,, H^S \} 
= 
c_4^\alpha\, \alpha + c_4^{\cal R}\, {\cal R} + c_4^{\cal E}\, {\cal E} 
 \approx 0
\,, 
\label{C4a}
\end{eqnarray}
where 
\begin{eqnarray}
c_4^\alpha &=& 
-\frac{8(\kappa_1-\kappa_3)}{\kappa_1^2 (2\kappa_1-3\kappa_3)^2} \left[
2k^4 \kappa_1^2 (2\kappa_1-\kappa_3)^2 
+ k^2 \kappa_1 \mu_1 
{(2\kappa_1-3\kappa_3)(2\kappa_1-\kappa_3)}
-6\mu_1^2(4\kappa_1^2-6\kappa_1\kappa_3 + 3\kappa_3^2)
\right]
\,,\\
c_4^{\cal R} &=& -
\frac{8(\kappa_1-\kappa_3)}{\kappa_1^2(2\kappa_1-3\kappa_3)}
\left[2k^4\kappa_1^2(2\kappa_1-\kappa_3)
+
k^2\kappa_1\mu_1
{(2\kappa_1-3\kappa_3)}
{-}6\mu_1^2(4\kappa_1-3\kappa_3)
\right]
\,, \\
c_4^{\cal E} &=& 
\frac{8(\kappa_1-\kappa_3)}{\kappa_1^2 (2\kappa_1-3\kappa_3)^2} \left[
2k^4 \kappa_1^2 (2\kappa_1-\kappa_3)^2 
{- k^2\kappa_1 \mu_1}
{(2\kappa_1+3\kappa_3)(2\kappa_1-\kappa_3)}
{-}2 \mu_1^2(2\kappa_1-3\kappa_3) (4\kappa_1-3\kappa_3)
\right]
\,.
\end{eqnarray}
The Poisson brackets between this constraint and primary constraints are 
	\begin{eqnarray}
	\{{\cal C}_4^\alpha, {\cal C}_1^\alpha\} 
	= - \frac{192 \mu_1^2 (\kappa_1 - \kappa_3)^2}{\kappa_1 (2 \kappa_1 - 3 \kappa_3)^2}
	\,, \qquad 
	\{{\cal C}_4^\alpha, {\cal C}_1^\beta\} = 0\,.
	\end{eqnarray}
Then, the consistency of this constraint $\dot{{\cal C}}^\alpha_4\approx 0$ fixes the Lagrange multiplier $\lambda_\alpha = -\{{\cal C}_4^\alpha, H^S\}/\{{\cal C}_4^\alpha, {\cal C}_1^\alpha\}$.
There are six second class constraints ${\cal C}_1^\alpha, {\cal C}_1^\beta, \widetilde{{\cal C}}_2^\alpha, {\cal C}_2^\beta, {\cal C}_3^\alpha,$ and ${\cal C}_4^\alpha$. Therefore, 
\begin{eqnarray}
\textrm{scalar DOF} = \frac{4\times 2 - 6 \, (\textrm{2 primary \& 2 secondary \& 1 tertiary \& 1 quaternary})}{2}
=1 \,.
\end{eqnarray}
In this case, the total number of degrees of freedom is five (Case V2).
It should be noted that the Fierz-Pauli theory is included in this class
since the linearized Einstein-Hilbert term satisfy the ``Condition 3'' \eqref{condition:3}, and the condition $\mu_1 = -\mu_2$ is included in the ``Condition 4'' \eqref{condition:4}.
Therefore, this is a wider class of Fierz-Pauli theory with five degrees of freedom for a massive spin-2 field.
The whole parameter family of this case can be mapped from Fierz-Pauli theory by a field redefinition as we will see in the next section.

\vspace{2mm}

{\bf Other case :} $\kappa_1 = \kappa_3$, $\kappa_4 = - (4 \kappa_1^2 - 4 \kappa_1 \kappa_3 + 3 \kappa_3^2)/(8 \kappa_1) = - (3/8) \kappa_1$ \\
Before the end of this section, we shall consider the case with $\kappa_1 = \kappa_3$
 and $\mu_1 = \mu_2 = 0$ or $\mu_1 = - 4 \mu_2 \neq 0$
 in which only one secondary constraint, ${\cal C}_2^\beta$ \eqref{const:c2beta}, exists
  while ${\cal C}_2^\alpha$ \eqref{const:c2alpha} vanishes
  and hence no more constraint will be generated as for ${\cal C}_2^\alpha$.

First the condition for the mass parameter in Case SIIc, ``Condition 4", is equivalent to 
\begin{eqnarray}
 \mu_2 = - \frac{4 \kappa_1^2-6 \kappa_1 \kappa_3+3 \kappa_3^2}{4 \kappa_1^2} \mu_1 
 = - \frac{1}{4}{\mu_1} \,.
\end{eqnarray}
Since $\{{\cal C}_2^\beta\,, {\cal C}_1^\beta\}\propto\mu_1\neq 0$, $\lambda_\beta$ can be determined by imposing $\dot{{\cal C}}^\beta_2\approx 0$,
that is  $\lambda_\beta = -\{{\cal C}_2^\beta, H^S\}/\{{\cal C}_2^\beta, {\cal C}_1^\beta\}$.
Since ${\cal C}_1^\alpha$ is first-class while ${\cal C}_{1,2}^\beta$ are second-class, we have
\begin{eqnarray}
\textrm{scalar DOFs} = \frac{4\times 2 - 1 \,(\textrm{1 primary})\times 2 \,(\textrm{first-class})
 - 2 \,(\textrm{1 primary \& 1 secondary})}{2}
=2 \,.
\end{eqnarray}
As we see in Appendix \ref{BDproof}, one of the modes is a ghost and hence we will no longer consider this case.

\vspace{2mm}

{\bf Case SW :} $\kappa_1 = \kappa_3$, $\kappa_4 = - (4 \kappa_1^2 - 4 \kappa_1 \kappa_3 + 3 \kappa_3^2)/(8 \kappa_1) = - (3/8) \kappa_1$, $\mu_1 = \mu_2 = 0$, \\
In this case, which is similar to Case SIIa except for the additional condition, $\kappa_1 = \kappa_3$,
 the time consistency of ${{\cal C}}^\beta_2$ yields a tertiary constraint 
 since $\{ {\cal C}_2^\beta \,, {\cal C}_1^\beta \}$ vanishes,
\begin{eqnarray}
 {\cal C}_3^\beta &\equiv & \dot{{\cal C}}_2^\beta
 = \{ {\cal C}^\beta_2 \,, H_T^S \} = \{ {\cal C}^\beta_2 \,, H^S \} 
 = 4 k^3 \kappa_1 (\alpha - {\cal R} - {\cal E}) \approx 0 \,.
\end{eqnarray}
The Poisson brackets with primary constraints are trivially satisfied and also the time-evolution of ${\cal C}_3^\beta$ turns out to be $\dot{{\cal C}}_3^\beta \propto {\cal C}_2^\beta$.
Since all the constraints commute each other, two primary, secondary, and tertiary constraints are first-class. Therefore
\begin{eqnarray}
\textrm{scalar DOF} = \frac{4\times 2 - 4 \,(\textrm{2 primary \& 1 secondary \& 1 tertiary})\times 2 \,(\textrm{first-class})}{2}
=0 \,.
\end{eqnarray}
The total number of degrees of freedom is two with 2 tensor modes and without vector mode corresponding to Case V1.

\section{Theoretical Properties}
\label{Sec:4}
In this section, we investigate theoretical properties of the obtained theories in the previous section in detail. We first consider a field redefinition linear in $h_{\mu \nu}$, which helps us to understand structures and classification of the theories. Furthermore, we clarify the crucial differences between the case SI and SIIb, which cannot be obtained from the known theories (linearized general relativity and Fierz-Pauli theory) 
by any invertible field redefinition.

\subsection{Linear field redefinition}
\label{subsec:LFD}
In this subsection, we consider a linear field redefinition of the rank-2 tensor $h_{\mu\nu}$, which respects Lorentz invariance.
A possible field redefinition as studied in \cite{Alvarez:2006uu,Bonifacio:2015rea} is
\begin{eqnarray}
h_{\mu\nu} = \Omega^2 \, {\overline h_{\mu\nu}} + \Gamma\, {\overline h} \, \eta_{\mu\nu} \,,
\label{eq:field redef}
\end{eqnarray}
where $\Omega$ and $\Gamma$ are constants, ${\overline h}$ is the trace of $\overline h_{\mu\nu}$ contracted by $\eta_{\mu\nu}$. Hereafter we set $\Omega=1$ without loss of generality since 
it only changes the normalization of the Lagrangian. 
The inverse transformation is given by 
\begin{eqnarray}
{\overline h}_{\mu\nu} = h_{\mu\nu} - \frac{\Gamma}{1+ 4\Gamma}\, h\, \eta_{\mu\nu} \,.
\end{eqnarray}
When $\Gamma = - 1/4$, this transformation is not invertible since its determinant vanishes.

Now we apply this transformation to our theories. 
After the field redefinition, the Lagrangian \eqref{Lagrangian} reads
\begin{align}
	S \Bigl[ h_{\mu \nu} = {\overline h_{\mu\nu}} + \Gamma\, {\overline h} \, \eta_{\mu\nu} \Bigr] 
	= \int {\rm d}^4 x \Bigl(-
	{\overline {\cal K}}^{\alpha \beta | \mu \nu \rho \sigma} {\bar h}_{\mu \nu, \alpha} {\bar h}_{\rho \sigma, \beta}
	-{\overline {\cal M}}^{\mu \nu \rho \sigma} {\bar h}_{\mu \nu} {\bar h}_{\rho \sigma} \Bigr) \,,
	\label{Lagrangian2}
\end{align}
where 
\begin{align}
	{\overline {\cal K}}^{\alpha \beta | \mu \nu \rho \sigma}
	&= {\overline \kappa}_1 \eta^{\alpha \beta} \eta^{\mu \rho} \eta^{\nu \sigma}
	+ {\overline \kappa}_2 \eta^{\mu \alpha} \eta^{\rho \beta} \eta^{\nu \sigma}
	+ {\overline \kappa}_3 \eta^{\alpha \mu} \eta^{\nu \beta} \eta^{\rho \sigma}
	+ {\overline \kappa}_4 \eta^{\alpha \beta} \eta^{\mu \nu} \eta^{\rho \sigma} \,,\\ 
	{\overline {\cal M}}^{\mu \nu \rho \sigma}
	&= \overline\mu_1 \eta^{\mu \rho} \eta^{\nu \sigma}
	+ \overline\mu_2 \eta^{\mu \nu} \eta^{\rho \sigma} \,.
\end{align}
The straightforward calculation shows the relation between $\kappa_i, {\mu_i}$ and ${\overline \kappa}_i, {{\overline\mu}_i}$ :
\begin{eqnarray}
&&{\overline \kappa}_1 = \kappa_1\,, \qquad 
{\overline \kappa}_2 = \kappa_2\,, \qquad 
{\overline \kappa}_3 = {2\Gamma\kappa_2+(1+4\Gamma)\kappa_3},\\ &&{\overline \kappa}_4 = 2\Gamma(1+2\Gamma)\kappa_1+\Gamma^2\kappa_2+\Gamma(1+4\Gamma)\kappa_3+(1+4\Gamma)^2\kappa_4\,,
\\ 
&& \overline\mu_1 = \mu_1\,, \qquad 
\overline\mu_2 = {2\mu_1\Gamma(1+2\Gamma)+\mu_2(1+4\Gamma)^2}
\,.
\end{eqnarray} 
It should be noted that coefficients $\kappa_{1,2}$ and $\mu_1$ are invariant under this transformation as long as $\Omega = 1$.
We also find that the degeneracy conditions for the scalar and vector modes, Eqs.~\eqref{condition:2} and \eqref{condition:3}, 
are invariant under the field redefinition Eq.~\eqref{eq:field redef}. 
Namely when $\kappa_i$ and $\mu_i$ satisfy the ``Condition 2'' and ``Condition 3'',
the transformed parameters still satisfy the same relations, that is
\begin{eqnarray}
	``\text{Condition 2}" : 2\overline\kappa_1+\overline\kappa_2=0 \,, \qquad
	``\text{Condition 3}" : 4\overline\kappa_1^2-4\overline\kappa_1\overline\kappa_3
	+3\overline\kappa_3^2
	+8\overline\kappa_1\overline\kappa_4 =0
	\,.
\end{eqnarray}
In addition to these, the additional condition to remove the extra degree of freedom for the scalar modes, $\mu_1=0$ for Case SI/SIIa\,(S${\overline{\rm IIa}}$)/SIIb\,(S${\overline{\rm IIb}}$) or
the ``Condition 4'' \eqref{condition:4} for Case SIIc, is also an invariant quantity.
This immediately leads that linear degenerate theories can be transformed to different linear degenerate theories
with a different parameter set through the field redefinition Eq.~\eqref{eq:field redef}. 
In particular, when one chooses
\begin{eqnarray}
	\Gamma ={-\frac{2\kappa_1-3\kappa_3}{12(\kappa_1-\kappa_3)}}
	\,,\label{TransG}
\end{eqnarray}
one can obtain the degenerate theories with $\overline\kappa_3=2\overline\kappa_1/3$ like in Case S${\overline{\rm IIa}}$ from the degenerate theories with $\kappa_3\neq 2\kappa_1/3$ like in Case SIIa.
We also note that the same kinetic structure as linearized 
Einstein-Hilbert term can be obtained for all the theories in Case SII by setting $\overline\kappa_3=2\overline\kappa_1$ via the field redefinition with
\begin{eqnarray}
	\Gamma =-\frac{2\kappa_1-\kappa_3}{4(\kappa_1-\kappa_3)}
	\,.\label{GammaToGR}
\end{eqnarray}
Hence, upon the field redefinition \eqref{eq:field redef}, one can show that 
\begin{eqnarray}
\textrm{Case SIIa and S${\overline{\rm IIa}}$} ~&\longleftrightarrow&~ \textrm{Linearlized general relativity} \,, \\
\textrm{Case SIIb} ~ &\longleftrightarrow& ~ \textrm{Case S$\overline{{\rm IIb}}$} \,, \\
\textrm{Case SIIc and S${\overline{\rm IIc}}$} ~ &\longleftrightarrow & ~ \textrm{Fierz-Pauli theory} \,.
\end{eqnarray}

\subsection{Symmetry of Lagrangian}
It is interesting to see the symmetry of Lagrangian
 as also studied in \cite{Alvarez:2006uu,Bonifacio:2015rea}. 
When $2 \kappa_1 + \kappa_2 = 0$ is satisfied, 
Lagrangian \eqref{reLagrangian} can be rewritten as assuming $\kappa_1 \neq \kappa_3$, 
\begin{align}
S [h_{\mu \nu}] 
= \kappa_1 S^\mathrm{LGR} [\widetilde{h}_{\mu \nu} ]
 &- \int {\rm d}^4 x \, \kappa_1 
 \frac{4 \kappa_1^2 - 4 \kappa_1 \kappa_3 + 3 \kappa_3^2 + 8 \kappa_1 \kappa_4}{8
 (\kappa_1 - \kappa_3)^2} \, \widetilde{h}_{, \alpha} \widetilde{h}^{, \alpha} \notag\\
 & - \int {\rm d}^4 x \left[ 
 \mu_1 \, \widetilde{h}_{\mu \nu} \widetilde{h}^{\mu \nu} 
+\frac{(2\kappa_1-\kappa_3)\kappa_3\mu_1+4\kappa_1^2\mu_2}{4(\kappa_1-\kappa_3)^2}\, 
\widetilde{h}^2 \right] \,,
\label{lagwithC1}
\end{align}
 where
\begin{align}
 \widetilde h_{\mu \nu} \equiv h_{\mu \nu} + \frac{\kappa_3- 2\kappa_1}{4\kappa_1} \, h \, \eta_{\mu \nu} \,,
\end{align}
and $S^\mathrm{LGR}$ stands for action of linearlized general relativity:
\begin{align}
 S^\mathrm{LGR} [h_{\mu \nu} ]
  \equiv - \int {\rm d}^4 x \Bigl( h_{\mu \nu \,, \alpha} h^{\mu \nu \,, \alpha}
 - 2 h^\alpha{}_{ \mu \,, \alpha} h^{\beta \mu}{}_{, \beta}
+ 2 h^{\alpha \beta}{}_{, \alpha} h_{, \beta}
- h_{, \alpha} h^{, \alpha} \Bigr) \,.
 \end{align}
First we should note that the kinetic term of tensorial parts in $\widetilde{h}_{\mu \nu}$
 is nicely summarized in a single term, that is $S^\mathrm{LGR}$.
 This is not possible with a generic $\kappa_2$.
The difference between $\widetilde{h}_{\mu \nu}$ and $h_{\mu \nu}$ is subtle
 since $\widetilde{h}_{\mu \nu}$ reduces to $h_{\mu \nu}$ once $\kappa_3 = 2 \kappa_1$ holds,
 which can be realized by a suitable field redefinition as studied in the last subsection.
 Hence its difference is not essential in discussion below at least in the absence of matter.

Now let us study the symmetry of Lagrangian. The first term $S^\mathrm{LGR}$ is invariant under 
a transformation with diffeomorphisms for $\widetilde{h}_{\mu \nu}$:
\begin{align}
 \widetilde{h}_{\mu \nu} \to \widetilde{h}_{\mu \nu} + \partial_\mu \xi_\nu + \partial_\nu \xi_\mu\,,
\end{align}
while other terms do not in general.
However, once the gauge parameters satisfy a condition such that $\partial_\mu \xi^\mu = 0$,
the kinetic term enjoys this restricted
gauge symmetry since $\widetilde{h}$ is invariant under this gauge transformation. 
Moreover, if $\mu_1$ also disappears, this symmetry becomes the symmetry of the whole Lagrangian.
This subgroup of the diffeomorphisms is known as transverse diffeomorphisms~\cite{Alvarez:2006uu}.

It is now clear that Case SI and SIIb (S$\overline{\rm IIb}$) possess such the transverse gauge symmetry
with $3$ gauge parameters, $1$ for scalar sector and $2$ for vector sector, 
since $\mu_1$ is vanishing as studied in detail in Appendix \ref{App:gauge}.
On the other hand, Case SIIa (S$\overline{\rm IIa}$) respects diffeomorphisms 
with $4$ gauge parameters since $\mu_1 = \mu_2 =0$ as in linearlized general relativity.
There is no gauge symmetry in Case SIIc (S$\overline{\rm IIc}$) since both $\mu_1$ and $\mu_2$ are present.
As for the remaining case SW, the Lagrangian can be written as 
\begin{align}
	S^\mathrm{SW} [h_{\mu \nu}]
	&= - \kappa_1 \int {\rm d}^4 x \left[ h_{\mu \nu \,, \alpha} h^{\mu \nu \,, \alpha}
	- 2 h^\alpha{}_{ \mu \,, \alpha} h^{\beta \mu}{}_{, \beta}
	+ h^{\alpha \beta}{}_{, \alpha} h_{, \beta}
	- \frac{3}{8} h_{, \alpha} h^{, \alpha} \right] \notag\\
	&= - \kappa_1 \int {\rm d}^4 x \Bigl[ \widehat{h}_{\mu \nu \,, \alpha} \widehat{h}^{\mu \nu \,, \alpha}
	- 2 \widehat{h}^\alpha{}_{ \mu \,, \alpha} \widehat{h}^{\beta \mu}{}_{, \beta} \Bigr] 
	= \kappa_1 S^\mathrm{LGR} 
	\Bigl[ \, \widehat{h}_{\mu \nu} 
	\equiv h_{\mu \nu} - \frac{1}{4} h \, \eta_{\mu \nu} \Bigr] \,.
\end{align}
Since the Lagrangian can be described in the form of lineralized general relativity
 with traceless tensor $\widehat{h}_{\mu \nu} \equiv h_{\mu \nu} - (1/4) h \, \eta_{\mu \nu}$,
  this theory enjoys diffeomorphism invariance as well as the invariance under a field redefinition \eqref{eq:field redef}.
 This enhanced symmetry is called as Weyl-invariant transverse diffeomorphism invariance \cite{Alvarez:2006uu}.

\subsection{New theories}
We have seen that the case SI, SIIb and S$\overline{\rm{IIb}}$ are the new theories of a spin-2 field, which cannot be mapped into linearized general relativity and Fierz-Pauli theory by the field redefinition as studied in the previous subsection \ref{subsec:LFD}. In this subsection, we take a closer look at the Lagrangian of these theories particularly focusing on its scalar sectors.
 
Now let us first consider linearized general relativity to understand the structure of theories, which can be simply obtained by setting $2\kappa_1 - \kappa_3 =0$ in the case SIIa or equivalently performing a field redefinition with a special $\Gamma$ \eqref{GammaToGR}. Rewriting the Lagrangian in the scalar sector in terms of the gauge invariant variables \eqref{GIVIIb}, we obtain up to total derivative terms
\begin{eqnarray}
{\cal L}^S_{\rm LGR} &=& -3 \dot{{\cal R}}^2 + k^2 {\cal R}^2 + 2k^2 \widetilde{\alpha} {\cal R}\,, \qquad
 \widetilde{\alpha} = \alpha + \frac{\dot{\beta}}{k} - \frac{\ddot{\cal E}}{k^2} \,, 
 \label{lag:LGR}
\end{eqnarray}
where we have set the overall factor $\kappa_1$ to be $1/8$ for simplicity. Then the variation with respect to $\widetilde{\alpha}$ gives the constraint ${\cal R}=0$, and it is manifest that the Lagrangian becomes zero after substituting the constraint. Thus we have confirmed that the number of degree of freedom in the scalar sector is zero in the Lagrangian formalism, that is consistent with the Hamiltonian analysis in the previous section.

\vspace{2mm}
	{\bf Case SIIb \& S$\overline{\rm \bf IIb}$}\ \ 
Now we would like to perform the same analysis for the case IIa. Again, we can set  $2\kappa_1 - \kappa_3 =0$ and  $\kappa_1 = 1/8$ without loss of generality. Then the Lagrangian in the scalar sector in terms of the gauge invariant variables \eqref{GIVI} is given by
\begin{eqnarray}
{\cal L}^S_{\rm IIb} &=& -3 \dot{{\cal R}}^2 + k^2 {\cal R}^2 + 2k^2 {\widetilde \alpha} {\cal R}
 -4\mu_2 \, {\widetilde {\cal E}}^2 \,, \qquad
 \widetilde{{\cal E}} = {\cal E} - \alpha - 3 {\cal R} \,.
\end{eqnarray}
One can clearly see that the first three terms are the exactly same as in the case of SIIa (S$\overline{\rm IIa}$). 
However, in this case, there is an extra term, $\mu_2\,{\widetilde {\cal E}}^2 =\mu_2({\rm Tr}\, h_{\mu\nu})^2$
 where $- \widetilde{{\cal E}}$ is the trace of a spin-2 field.
Since it is completely decoupled from ${\cal R}$ and $\widetilde{\alpha}$ and has no kinetic term,  it is actually a non-dynamical degree of freedom.
 The case SIIb (S$\overline{\rm IIb}$) cannot be obtained from any gauge fixing of the case SIIa (S$\overline{\rm IIa}$), and these theories are therefore independent each other.  Since $\widetilde{\alpha}$ and $\widetilde{{\cal E}}$ are non-dynamical, the Lagrangian becomes zero after integrating out these variables just as in the case SIIa (S$\overline{\rm IIa}$).

\vspace{2mm}
	{\bf Case SI}\ \ 
Finally, we consider the case SI, whose number of scalar degree of freedom is one. 
In this case, we can also choose $2\kappa_1 - \kappa_3 =0$ and $\kappa_1 = 1/8$ without loss of generality.
Furthermore, we set $\kappa_4=1/8$ to simplify the coefficient of the kinetic term for $\widetilde{{\cal E}}$. 
Then the Lagrangian in terms of the gauge invariant variables \eqref{GIVI} is given by
\begin{eqnarray}
{\cal L}^S_{\rm I} &=& -3 \dot{{\cal R}}^2 + k^2 {\cal R}^2 + 2k^2 {\widetilde \alpha} {\cal R} 
+\dot{\widetilde {\cal E}}{}^2-(k^2+4\mu_2) {\widetilde {\cal E}}^2 \,.
\end{eqnarray}
In this case, the trace of $h_{\mu\nu}$, namely $- \widetilde{\cal E}$, which again decouples from other variables ${\cal R}$ and $\widetilde \alpha$, becomes dynamical since it has now the kinetic term, differently from the case SIIb. Thus after plugging the constraint of $\widetilde \alpha$, one scalar degree of freedom remains in the case SI, which is also consistent with the Hamiltonian analysis.

\section{Summary \& Discussion}
\label{sec:summary}

\begin{table}[!t]
	\begin{center}
		\begin{tabular}{|c||c|c|c|c|} \hline
			Case & DOF  & Conditions & Free parameters & Comments \\ \hline \hline
			SI \& V1 & ~$3=2+0+1$~ & $\mu_1 = 0$  & $\kappa_3, \kappa_4, \mu_2$ & New theories \\
			\hline 
			SIIa \& V1 & $2=2+0+0$ &  ``Condition 3'' \& $\mu_1=\mu_2=0$   & $\kappa_3$ & General relativity is included \\
			\hline 
			SIIb \& V1 & $2=2+0+0$ &   ``Condition 3'' \& $\mu_1 = 0$ \& $\mu_2\neq 0$ & $\kappa_3, \mu_2$ & New theories  \\
			\hline
			SIIc \& V2 & $5=2+2+1$ &  ``Condition 3 \& 4" \& $\mu_1 \neq 0$  & $\kappa_3, \mu_1$ & Fierz-Pauli is included\\ 
			\hline\hline
			S$\overline{\rm{IIa}}$ \& V1 & $2=2+0+0$ & ``Condition 5'' \& $\mu_1=\mu_2=0$   & {\rm None}&  ~~$-2\kappa_1 + 3\kappa_3 = 0$ limit of  SIIa~~\\ 
			\hline	
			S$\overline{\rm{IIb}}$ \& V1 & $2=2+0+0$ &  ``Condition 5'' \& $\mu_1 = 0$ \& $\mu_2\neq 0$  & $\mu_2$ & ~~$-2\kappa_1 + 3\kappa_3 = 0$ limit of SIIb~~  \\ 
			\hline
			S$\overline{\rm{IIc}}$ \& V2 & $5=2+2+1$ & ``Condition 5'' \& $\mu_1 + 3\mu_2= 0$ \& $\mu_1 \neq 0$ & $\mu_1$ & $-2\kappa_1 + 3\kappa_3 = 0$ limit of SIIc \\ 
			\hline
		\end{tabular}
		\caption{The number of the degrees of freedom, the conditions, free parameters and comments for each case is shown. For any case, the ``Condition 1'' (\ref{condition:1}) and  the ``Condition 2'' (\ref{condition:2}) 
			are always imposed. ``Condition 3,4,5'' are shown in \eqref{condition:3}, \eqref{condition:4}, and \eqref{condition:5} respectively. Among free parameters, $\kappa_1$ is not included since it only changes the normalization of the Lagrangian if its sign is appropriately chosen.}
		\label{table}
	\end{center}
\end{table}

{\bf Summary }
In the present paper, we constructed the most generic spin-2 field theories in a flat spacetime
with at most five degrees of freedom without ghost mode, whose Lagrangian consists of the quadratic terms of the field and its first derivatives. 
By decomposing the spin-2 field $h_{\mu\nu}$ into the transverse-traceless tensor, tensor composed by transverse vectors, and tensor composed by scalar components, we classified theoretically consistent theories based on the Hamiltonian analysis in a systematic manner. 

We found that the existence of the tensor degrees of freedom is always controlled by one parameter $\kappa_1$, which is assumed to be non-zero while we imposed the degeneracy conditions in order to eliminate extra problematic degrees of freedom for the vector and scalar modes. 
Under the degeneracy conditions, 
we found two classes in the vector sector : 2 propagating vector degrees of freedom and no degrees of freedom. As in the vector sector, we have also classified theories in the scalar sector based on the Hamiltonian analysis. 
The classification of the obtained theories is summarized in Table.~\ref{table}. The case SIIa and SIIc are a wider class of the known theories : linearized general relativity and Fierz-Pauli theory, and we have shown that the case SIIa and SIIc can be mapped from these known theories by field redefinition. On the other hand, the case SI and SIIb are new theories, which cannot be mapped from the known theories. The case SIIb has the same number of degrees of freedom in linearized general relativity, however, it has less gauge degrees of freedom, and it contains the trace of the spin-2 field, which is non-dynamical. On the other hand, the case I has the dynamical degree of freedom coming from the trace of the spin-2 field while the vector degrees of freedom is absent. The other remaining three cases: S$\overline{\rm IIa}$, S$\overline{\rm IIb}$, and S$\overline{\rm IIc}$, are the subset of the cases SIIa, SIIb, and SIIc and hence the case S$\overline{\rm IIb}$ is also a new theory. We provided the conditions for avoiding the ghost, gradient, and tachyonic instabilities by calculating reduced Lagrangian in Appendix \ref{App:gauge}.

\vspace{2mm}
{\bf Non-linear extension to gravity theory}
It is interesting to consider a non-linear extension to (massive) gravity theory of the obtained theories for a spin-2 field. Let us consider the following 
Lagrangian\footnote{
The effect of the background curvature in the Lagrangian has been discussed in the literature, e.g. \cite{Akagi:2018ufw}, which is beyond the scope of the present paper.}, 
\begin{eqnarray}
S = \int {\rm d}^4x \sqrt{-g}
\left[
a_ 1 R + a_2 hR + a_3 h_{\mu\nu} R^{\mu\nu}  + 
{\mathfrak K}^{\alpha \beta | \mu \nu \rho \sigma} \nabla_\alpha h_{\mu \nu} \nabla_\beta h_{\rho \sigma}
+
{\mathfrak M}^{\mu \nu \rho \sigma} h_{\mu \nu} h_{\rho \sigma}  + {\cal O} (h^2R, h^3)
\right] \,,
\label{MGaction}
\end{eqnarray}
where the fluctuation tensor is defined by $h_{\mu\nu} = g_{\mu\nu} -\eta_{\mu\nu}$. The coefficients of the kinetic and mass terms, ${\mathfrak K}^{\alpha \beta | \mu \nu \rho \sigma}$ and $
{\mathfrak M}^{\mu \nu \rho \sigma}$, are functions of the metric
$g_{\mu\nu}$, 
\begin{align}
{\mathfrak K}^{\alpha \beta | \mu \nu \rho \sigma}
&= b_1 g^{\alpha \beta} g^{\mu \rho} g^{\nu \sigma}
+ b_2 g^{\mu \alpha} g^{\rho \beta} g^{\nu \sigma}
+ b_3 g^{\alpha \mu} g^{\nu \beta} g^{\rho \sigma}
+ b_4 g^{\alpha \beta} g^{\mu \nu} g^{\rho \sigma} \,,\\ 
{\mathfrak M}^{\mu \nu \rho \sigma}
&= \mu_1 g^{\mu \rho} g^{\nu \sigma}
+ \mu_2 g^{\mu \nu} g^{\rho \sigma} \,,
\end{align}
and $a_{1,2,3}$ and $b_{1,2,3,4}$ are constant parameters.
By expanding the action up to the quadratic terms in $h_{\mu\nu}$ and identifying this field as the (massive) spin-2 field introduced in this paper, 
one can find that the relations between these constant parameters and the $\kappa$ parameters introduced in the Lagrangian \eqref{Lagrangian} are given by
\begin{eqnarray}
\kappa_1 = b_1 -{1 \over 2} a_1 + {1 \over 2}a_3 \,, \qquad
\kappa_2 = b_2 + a_1 -a_3 \,, \qquad
\kappa_3 = b_3 -a_1 -a_2 -{1 \over 2}a_3 \,, \qquad
\kappa_4 = b_4 + {1 \over 2}a_1 + a_2 \,.
\end{eqnarray}
Then the gravitational action \eqref{MGaction} perturbed around Minkowski space-time reproduces the Lagrangian \eqref{Lagrangian} at the linear level. Therefore it is for sure that this Lagrangian \eqref{MGaction} with the conditions that we have obtained does not have an extra degrees of freedom at linear order. 
However, one might need some extra tuning of the parameters $a_i$ and $b_i$ for avoiding the appearance of Boulware-Daser ghost at nonlinear level, and it has to be carefully examined just as in the construction of the dRGT theory.

\vspace{2mm}
{\bf Matter coupling }
Although we have obtained the interesting theories of the spin-2 field, 
one carefully needs to introduce a coupling to matter fields. 
To clarify this, let us consider the case SIIc with the matter coupling $h_{\mu\nu} T^{\mu\nu}/M$, where $T^{\mu\nu}$ is the energy-momentum tensor of an external source, and $M$ is a mass parameter, which corresponds to the Planck's mass in general relativity. 
Ignoring the vector modes\footnote{A careful analysis shows that the decoupling limit Lagrangian of the vector field becomes massless $U(1)$ field, which completely decouples from other modes. },  
the $\Lambda_3$ decoupling limit with the Stuckelberg decomposition, $h_{\mu\nu}  = {\widehat h}_{\mu\nu} +2m^2 \partial_\mu \partial_\nu \pi +b \,m^2\,  \eta_{\mu\nu}\square \pi$, and the conformal transformation, 
${\widehat h}_{\mu\nu} = {\frak h}_{\mu\nu} - ({c_1}/{\kappa_1-\kappa_3}) \pi \eta_{\mu\nu}$
yields the following Lagrangian, 
\begin{eqnarray}
{\cal L}^{\rm (DL)}&=&
{\cal L}^{\rm (DL)}_{\rm tensor}[{\frak h}]
-  {6c_1 \over \kappa_1} (\partial_\mu \pi)^2 
+ {1 \over M} \left( {\frak h}_{\mu\nu} T^{\mu\nu}  
- \frac{c_1}{\kappa_1-\kappa_3}\pi T\right)
+ {1 \over \Lambda_3^3} \Bigl[ 2 (\partial_\mu \partial_\nu \pi) T^{\mu\nu} + b (\square \pi) T \Bigr]\,, 
\end{eqnarray}
where ${\cal L}^{\rm (DL)}_{\rm tensor}[{\widetilde h}]$ is the kinetic Lagrangian of 
the case IIc, $\mu_1 = c_1 m^2$, $m$ is a mass parameter, $T=\eta_{\mu\nu} T^{\mu\nu}$
and $\Lambda_3 = (M m^2)^{1/3}$.
Here, the new interaction appears if $b\neq 0$ and $T \neq 0$
which cannot be appeared in the Fierz-Pauli theory with minimally coupled matter. 
The equation of motion for the scalar mode $\pi$ naively contains 
the second derivatives of the energy-momentum tensor, which might lead
to the higher derivatives, depending on matter fields. 
Whether this matter coupling introduce an extra degree of freedom associated with higher derivatives or not will be reported in future work (see also \cite{Arnowitt:1962hi}).

\acknowledgments
We would like to thank Claudia de Rham, Kazuya Koyama and Nobuyoshi Ohta for fruitful discussions and useful suggestions.
This work was supported in part by JSPS Grant-in-Aid for Scientific Research 
Nos.~JP17K14276 (RK), JP25287054 (RK), 17K14304 (DY). 
A.N. is supported in part by a JST grant ``Establishing a Consortium for the Development of Human Resources in Science and Technology''. The work of A.N. is also partly supported by the JSPS Grant-in-Aid for Scientific Research No.16H01092.

\appendix

\section{Other degenerate theories in the scalar sector}
\label{sec:other cases}

In this Appendix we consider special cases with $2\kappa_1 - 3\kappa_3 = 0$, 
\begin{eqnarray}
 ``\text{Condition 5}" \ :\ \kappa_3 = \frac{2}{3} \kappa_1 \,, 
\label{condition:5}
\end{eqnarray}
which also satisfy ``Condition $3$" \eqref{condition:3} as well as ``Condition $1$"  \eqref{condition:1} 
\& ``Condition $2$" \eqref{condition:2}. 
To summarize, we have
\begin{eqnarray}
 \kappa_1 > 0 \,, \qquad
 \kappa_2 = -2\kappa_1 \,, \qquad
 \kappa_3 = \frac{2}{3} \kappa_1 \,, \qquad
 \kappa_4 = - \frac{4 \kappa_1^2 - 4 \kappa_1 \kappa_3 + 3 \kappa_3^2}{8 \kappa_1} 
 = - \frac{\kappa_1}{3} \,.
\end{eqnarray}
In this case, the momenta are given by
\begin{eqnarray}
\pi_\alpha = {16 \over 3} \kappa_1 (k \beta -\dot{\alpha}) \,, \qquad
\pi_{\cal E} = -{16 \over 3} \kappa_1 (k \beta -\dot{{\cal E}}) \,,
\end{eqnarray}
and other momenta gives two primary constraints, 
\begin{eqnarray}
{\cal C}_1^\beta = \pi_\beta =0 \,, \qquad {\cal C}_1^{\cal R}=\pi_{\cal R}=0 \,.
\label{primarybetacalR}
\end{eqnarray}
The Hamiltonian and the total Hamiltonian densities are given by 
\begin{align}
{\cal H}^S &= \dot{\alpha} \pi_\alpha + \dot{\beta} \pi_\beta + \dot{{\cal R}} \pi_{\cal R} + \dot{{\cal E}} \pi_{\cal E} - {\cal L}^S  \,, \nonumber \nonumber\\
&\approx 4\left[{2 k^2 \kappa_1 \over 3}+ (\mu_1+\mu_2)\right]\alpha^2
-2\mu_1\beta^2-4\left[{2 k^2 \kappa_1 \over 3}-(\mu_1+\mu_2)\right]{\cal E}^2
+12(\mu_1+3\mu_2) {\cal R}^2 \nonumber\\
&~~~
-{8 \over 3} (3\mu_1 + 9\mu_2 -2k^2\kappa_1) {\cal E} {\cal R}
- 8\left[\mu_2 {\cal E} + {1\over 3} (2k^2 \kappa_1 -9\mu_2) {\cal R}\right]\alpha-{3 \over 32\kappa_1} (\pi_\alpha^2 -\pi_{\cal E}^2) + {k} (\pi_\alpha + \pi_{\cal E}) \beta 
\,,  \\
{\cal H}_T^S&
= {\cal H}^S + \lambda_\beta \pi_\beta + \lambda_{\cal R} \pi_{\cal R} \,.
\end{align}
where we have suppressed the terms proportional to the primary constraints (\ref{primarybetacalR}) in the final expression of ${\cal H}^S$.
Then one finds
\begin{align}
{\cal C}_2^{{\cal R}} &\equiv \dot{{\cal C}}_1^{{\cal R}} = \{ {\cal C}_1^{{\cal R}} \,, H_T^S \} = \{ {\cal C}_1^{{\cal R}} \,, H^S \}
= -{8\over 3} \left[
(9\mu_2-2k^2\kappa_1)\alpha + (2k^2 \kappa_1 -3\mu_1 -9\mu_2){\cal E} +9(\mu_1 + 3\mu_2){\cal R}
\right]\\
{\cal C}_2^{\beta} &= \dot{{\cal C}}_1^{\beta} = \{ {\cal C}_1^{\beta} \,, H_T^S \} = \{ {\cal C}_1^{\beta} \,, H^S \}
=  -k\pi_\alpha -k\pi_{\cal E} +4\mu_1\beta \,.
\end{align}
The Poisson brackets between the secondary and the primary constraints are given by 
\begin{align}
\{{\cal C}_2^\beta, {\cal C}_1^\beta\} &= 4\mu_1 \,, \label{C2bC1b2}\\
\{{\cal C}_2^{\cal R}, {\cal C}_1^{\cal R}\} &= -24(\mu_1+3\mu_2) \,.\label{C2RC1R2}
\end{align}
If both are non-vanishing, the Lagrange multipliers $\lambda_\beta$ and $\lambda_{\cal R}$ are determined, 
then it has two primary constraints ${\cal C}_1^\beta, {\cal C}_1^{\cal R}$ and two secondary constraints ${\cal C}_2^\beta, {\cal C}_2^{\cal R}$. 
Since all the constraints are second-class, 
the total number of degrees of freedom in the scalar sector is $(8-4)/2=2$.
As shown in Appendix \ref{BDproof}, we have a ghost degree of freedom in this case.
To remove an extra degree of freedom, we have three options : 
$\mu_1=\mu_2=0$ ({\bf Case S$\overline{\rm \bf IIa}$}), $\mu_1=0$ ({\bf Case S$\overline{\rm \bf IIb}$}),
 or $\mu_1 + 3\mu_2=0$ ({\bf Case S$\overline{\rm \bf IIc}$}).

\vspace{2mm}
{\bf Case S${\overline {\rm \bf IIa}}$ :} 
$\mu_1=\mu_2=0$, 
\\
If $\mu_1=\mu_2=0$, both \eqref{C2bC1b2} and \eqref{C2RC1R2} vanish. 
In this case, all the primary constraints ${\cal C}_1^\beta, {\cal C}_1^{\cal R}$ and secondary constraints ${\cal C}_2^\beta, {\cal C}_2^{\cal R}$ are first-class since all the constraints commute with each other. 
Therefore,
\begin{eqnarray}
\textrm{scalar DOF} = \frac{4\times 2 - 4 \,(\textrm{2 primary \& 2 secondary})\times 2 \,(\textrm{first-class})}{2}
=0 \,.
\end{eqnarray}
Since the vector mode does not have a degree of freedom (Case V1), only the tensor mode propagate. This case corresponds to the case SIIb with an additional condition $-2\kappa_1 + 3\kappa_3 =0$ 
as one can see from the structure of the constraint algebra as well as Lagrangian more explicitly.

\vspace{2mm}

{\bf Case S${\overline {\rm \bf IIb}}$ :} 
$\mu_1 = 0$ \\
If $\mu_1=0$, the Lagrange multiplier $\lambda_{\cal R}$ can be determined by
$\dot{{\cal C}}_2^{\cal R} \approx 0$, that is, $\lambda_{\cal R} = -\{{\cal C}_2^{\cal R}, H^S\}/\{{\cal C}_2^{\cal R}, {\cal C}_1^{\cal R}\}$.
On the other hand, the consistency of ${\cal C}_2^\beta$ generates a tertiary constraint
\begin{eqnarray}
{\cal C}_3^\beta &\equiv & \dot{{\cal C}}_2^\beta 
= \{ {\cal C}^\beta \,, H_T^S \} = \{ {\cal C}^\beta \,, H^S \} =
{16 \over 3} k^3 \kappa_1 (\alpha-{\cal E})
\approx 0 \,.
\end{eqnarray}
One also finds $\dot{{\cal C}}_3^\beta \propto {\cal C}_2^\beta$ and hence no more constraint is generated.
${\cal C}_1^\beta$, ${\cal C}_2^\beta$, and ${\cal C}_3^\beta$ are first-class since these constrains commute with all other constraints, i.e., $\{{\cal C}_i^\beta, {\cal C}_j^\beta\}=\{{\cal C}_i^\beta, {\cal C}_k^{\cal R}\}=0$ where $i, j=1, 2, 3$ and $k = 1, 2$ while ${\cal C}_1^{\cal R}$ and ${\cal C}_2^{\cal R}$ are second-class. 
Therefore,
\begin{eqnarray}
\textrm{scalar DOF} = \frac{4\times 2 - 3 \, (\textrm{1 primary \& 1 secondary \& 1 tertiary})\times 2\, (\textrm{first-class}) - 2\,(\textrm{1 primary \& 1 secondary})}{2}
=0\,. \nonumber\\
\end{eqnarray}
Then the total number of degrees of freedom is two
 since the vector mode does not propagate for $\mu_1=0$ (Case V1) while tensor modes are present.

\vspace{2mm}

{\bf Case S${\overline {\rm \bf IIc}}$ :} 
$\mu_1 + 3\mu_2= 0$, $\mu_1 \neq 0$ \\
If $\mu_1 + 3\mu_2=0$, the Lagrange multiplier $\lambda_\beta$ can be determined by $\dot{{\cal C}}_2^\beta \approx 0$, i.e., $\lambda_\beta = -\{{\cal C}_2^\beta, H^S\}/\{{\cal C}_2^\beta, {\cal C}_1^\beta\}$. 
As for ${\cal C}_2^{\cal R}$, since it does not commute with ${\cal C}_2^\beta$, 
 we shall consider a linear combination of constraints:
\begin{eqnarray}
\widetilde{{\cal C}}_2^{\cal R} &\equiv & {\cal C}_2^{\cal R} - 2 k \, {\cal C}_1^\beta \,.
\end{eqnarray}
The consistency of $\widetilde{{\cal C}}_2^{\cal R}$ generates a tertiary constraint
\begin{eqnarray}
{\cal C}_3^{\cal R} &\equiv & 
\dot{\widetilde{{\cal C}}}{}_2^{\cal R} 
= \{ \widetilde{{\cal C}}_2^{\cal R} \,, H_T^S \} = \{ \widetilde{{\cal C}}_2^{\cal R} \,, H^S \} 
= \left( k^2 - {3\mu_1 \over 2\kappa_1} \right) \pi_\alpha + k^2 \pi_{\cal E}
\approx 0 \,.
\end{eqnarray}
Since the Poisson brackets between this constraint and primary constraints are vanishing
the time-evolution of the tertiary constraint ${\cal C}_3^{\cal R}$ yields a quaternary constraint
and in fact, 
\begin{eqnarray}
{\cal C}_4^{\cal R} &\equiv & 
\dot{{\cal C}}_3^{\cal R} 
= \{ {\cal C}_3^{\cal R} \,, H_T^S \} 
= \{ {\cal C}_3^{\cal R} \,, H^S \} 
\nonumber\\
&=&
 - 8 \left( \frac{2 \kappa_1 k^4}{3} - \frac{\mu_1^2}{\kappa_1} \right) \alpha 
 - \frac{12 \mu_1^2}{\kappa_1} {\cal R}
 + 4 \left(\frac{4 \kappa_1 k^4}{3} - 2 k^2 \mu_1 + \frac{\mu_1^2}{\kappa_1} \right) {\cal E}
\approx 0 \,.
\end{eqnarray}
Since this does not commute with ${\cal C}_1^{\cal R}$, no more constraint is generated and
 the time-consistency of ${\cal C}_4^{\cal R}$ determines the Lagrange multiplier as $\lambda_{\cal R} = - \{ {\cal C}_4^{\cal R} \,, H^S\}/\{ {\cal C}_4^{\cal R} \,, {\cal C}_1^{\cal R}\}$.
There are six second-class constraints ${\cal C}_1^{\cal R}, {\cal C}_1^\beta, \widetilde{{\cal C}}_2^{\cal R}, {\cal C}_2^\beta, {\cal C}_3^{\cal R},$ and ${\cal C}_4^{\cal R}$. Therefore,
\begin{eqnarray}
\textrm{scalar DOF} = \frac{4\times 2 - 6\,(\textrm{2 primary \& 2 secondary \& 1 tertiary \& 1 quaternary })}{2}
=1\,.
\end{eqnarray}
In this case, the total number of degrees of freedom is five combining two vector modes (Case V2)
 and two tensor modes. 
One may notice that the structure of the constraints and the number of the degrees of freedom are the same as the case SIIc. 
This implies that this case S${\overline{\rm IIc}}$ is the special case of the case SIIc with $-2\kappa_1 +3 \kappa_3 =0$.

\section{Appearance of unwanted ghost}
\label{BDproof}
In this appendix, we show that there is ghost if a system has two or more degrees of freedom in the scalar sector. More clearly, we will show that at least one of ghost modes originates from the structure of the Einstein-Hilbert action. Due to the general covariance of the Einstein-Hilbert action, the signs of kinetic terms for scalar and tensor modes are opposite. Assuming the Newton's constant is positive, the kinetic term for a scalar mode is wrong, as found in \eqref{lag:LGR}. In the case of general relativity, thanks to diffeomorphism invariance, there is no intrinsic degree of freedom in scalar sector and hence this ghost mode is not activated. But once the diffeomorphism invariance is lost, this ghost revives in general, which is explicitly shown below. 

The conditions to have two or more degrees of freedom are summarized as follows:
\begin{align}
 3 \, \mathrm{d.o.f.} &:  
 4 \kappa_1^2 - 4 \kappa_1 \kappa_3 + 3 \kappa_3^2 + 8 \kappa_1 \kappa_4 \neq 0 \,, \notag\\
 & \quad \mu_1 \neq 0 \,, \\
 2 \, \mathrm{d.o.f.} &:  
 4 \kappa_1^2 - 4 \kappa_1 \kappa_3 + 3 \kappa_3^2 + 8 \kappa_1 \kappa_4 = 0 \,, \notag\\
 & \quad \mu_1 \neq 0 \,, \notag\\
 & \quad
 \mu_1 (4 \kappa_1^2 - 6 \kappa_1 \kappa_3 + 3 \kappa_3^2) + 4 \mu_2 \kappa_1^2 \neq 0 \,, \label{secondcase}\\ 
 2 \, \mathrm{d.o.f.} &: 
 3 \kappa_1 = 3 \kappa_3 = - 8 \kappa_4 \,, \notag\\
 & \quad 
 \mu_1 = - 4 \mu_2 \neq 0 \,.
 \label{lastcase}
 \end{align}
In addition to these conditions, we further assume $2 \kappa_1 + \kappa_2 = 0$ and $\kappa_1 \neq 0$
 to evade ghost in the vector sector.
First we focus on the first two cases.
Then we find the basic structure of Lagrangian in the last case is essentially the same as in the first two cases and conclude the existence of ghost in the last case too. 
It should be noticed that we have 
 $4 \kappa_1^2 - 4 \kappa_1 \kappa_3 + 3 \kappa_3^2 + 8 \kappa_1 \kappa_4 = \kappa_1 (3 \kappa_1 + 8 \kappa_4)$ and $\mu_1 (4 \kappa_1^2 - 6 \kappa_1 \kappa_3 + 3 \kappa_3^2) + 4 \mu_2 \kappa_1^2
 = \kappa_1^2 (\mu_1 + 4 \mu_2)$ respectively thanks to the first condition in \eqref{lastcase}.

In order to simplify discussion, 
 we further assume $\kappa_3 =2\kappa_1$, 
 which is always realized by performing a field redefinition as studied in \ref{subsec:LFD}.
In order to show the existence of ghost, we shall take a look at Lagrangian \eqref{SS}, which now reduces to 
\begin{align}
 {\cal L}^S{[\alpha, \beta, {\cal R}, {\cal E}]} 
  &= 8 \kappa_1 \bigl( - 3 \dot{{\cal R}}^2 + k^2 {\cal R}^2 + 2 k^2 {\cal R} \widetilde{\alpha} \bigr)
 + 4 (\kappa_1 + \kappa_4) \bigl( \dot{\widetilde{{\cal E}}}{}^2 - k^2 \widetilde{{\cal E}}^2 \bigr) \notag\\
 & \qquad + 2 \mu_1 \bigl( - 4 \alpha^2 + \beta^2 - 12 {\cal R}^2 
 - 4 \alpha \widetilde{{\cal E}} - 8 \alpha {\cal R} - 8 {\cal R} \widetilde{{\cal E}} \bigr) - 4 (\mu_1 + \mu_2) \widetilde{{\cal E}}^2 \,.
 \label{lagghost}
\end{align}
Here, $\widetilde{\alpha}$ and $\widetilde{{\cal E}}$ stand for 
\begin{align}
 \widetilde{\alpha} = \alpha + \frac{\dot{\beta}}{k} - \frac{\ddot{\cal E}}{k^2} \,, \qquad
 \widetilde{{\cal E}} = - \mathrm{tr} [h_{\mu \nu}] = {\cal E} - \alpha - 3 {\cal R} \,.
\end{align}
Thanks to the condition $2 \kappa_1 + \kappa_2 = 0$, it is clear that Lagrangian consists of three parts, namely linearized general relativity (the first three terms), the kinetic term for
 the trace of $h_{\mu \nu}$, $- \widetilde{{\cal E}}$, and the mass terms as shown in \eqref{LS mass term}. 
A crucial term for the existence of ghost is the one proportional to $\ddot{{\cal E}}$ in $\widetilde{\alpha}$,
 which gives a cross term $\dot{{\cal R}} \dot{{\cal E}}$ after integration by part.
In the presence of $\mu_1$, the gauge invariance of the Lagrangian is totally lost and hence 
 ${\cal R}$, ${\cal E}$, and also possibly $\widetilde{{\cal E}} \sim \alpha$ are physical degrees of freedom
  while $\beta$ is non-dynamical since it has no kinetic term.
If $\kappa_1 + {\kappa_4}$ vanishes, the kinetic term for $\widetilde{{\cal E}}$ is lost, which ends up with two degrees of freedom.
Since $\beta$ is not a dynamical degree of freedom in any case, the constraint equation for $\beta$ yields
\begin{align}
 \beta = {\frac{4 k \kappa_1}{\mu_1}\dot{{\cal R}}}\,. 
\end{align}  
Then after eliminating the dependence of $\beta$ in the Lagrangian \eqref{lagghost}, the coefficient of $\dot{{\cal R}}^2$ can be changed.
But still one finds
\begin{align}
 {\cal L}{^S [{\cal R}, {\cal E}, \widetilde{{\cal E}}]}
 &= - {8 r \kappa_1} \dot{{\cal R}}^2 + 16 \kappa_1 \dot{{\cal R}} \dot{{\cal E}}
 + {4} (\kappa_1 + {\kappa_4}) \dot{\widetilde{{\cal E}}}{}^2 + {[\textrm{no time derivative terms}]}\notag\\
 &= - {8 r \kappa_1} \left( \dot{{\cal R}} - \frac{{1}}{r} \dot{{\cal E}} \right)^2 
 + \frac{8}{r} \kappa_1\dot{{\cal E}}^2 + {4}(\kappa_1 + {\kappa_4}) \dot{\widetilde{{\cal E}}}{}^2 + {[\textrm{no time derivative terms}]}\,, 
 \label{Lextra}
\end{align} 
where $r=3 + 4\kappa_1k^2/\mu_1$.
Now it is clear that no matter the signature of $r$ is, the kinetic terms for ${\cal R}$ and ${\cal E}$ have opposite sign.
Then at least either ${\cal R}$ or ${\cal E}$ will be a ghosty mode while the nature of $\widetilde{{\cal E}} \sim \alpha$ will be determined by the coefficient of $\kappa_1 + \kappa_4$, that is it is ghost or not or even non-dynamical. 
For \eqref{secondcase}, i.e., $\kappa_1+\kappa_4=0$, the appearance of a ghost mode is still inevitable
 since $\widetilde{{\cal E}} \sim \alpha$ constraint does not involve any time derivatives and the kinetic part of the Lagrangian remains the same as  \eqref{Lextra}.

Finally we shall consider the last case \eqref{lastcase}, whose Lagrangian reads
\begin{align}
 {\cal L}^S = 8 \kappa_1 \left[ - 3 \dot{\widetilde{{\cal R}}}{}^2 - 5 k^2 \widetilde{{\cal R}}^2 
+ 2 k^2 \widetilde{{\cal R}} \left( {\cal E} + \frac{\dot{\beta}}{k} - \frac{\ddot{{\cal E}}}{k^2} \right) \right] 
 + 2 \mu_1 \bigl( \beta^2 - 4 {\cal E}^2 - 24 \widetilde{{\cal R}}^2+ 16 {\cal E} \widetilde{{\cal R}} \bigr) \,,
\end{align} 
 where
\begin{align}
 \widetilde{{\cal R}} \equiv \frac{1}{4} (-\alpha + {\cal R} + {\cal E}) \,.
\end{align} 
We note that the essential structure of Lagrangian is quite similar to that in the cases studied above,  
\eqref{lagghost}. Due to the cross term $\widetilde{{\cal R}} \ddot{{\cal E}} \sim \dot{\widetilde{{\cal R}}} \dot{{\cal E}}$,
 one will find a ghost in a similar manner while the constraint equation for $\beta$ yields $\beta = (4 k \kappa_1/\mu_1) \dot{\widetilde{{\cal R}}}$.

\section{Gauge transformations and reduced Lagrangians}
\label{App:gauge}
In this Appendix, we derive a gauge transformation for the cases which have first-class constraints. By using gauge invariant quantities, we derive reduced Lagrangians and the conditions for avoiding ghost and gradient instabilities. The details of finding gauge transformations from a system with first-class constraints 
can be referred to \cite{Castellani1982,Sugano1982,Sugano1986,Sugano:1989rq}. 
Since the case S${\overline{\rm IIa}}$, S${\overline{\rm IIb}}$, and S${\overline{\rm IIc}}$ are  equivalent to SIIa, SIIb, and SIIc modulo the invertible field transformation (\ref{eq:field redef}), we will omit the detailed analysis for those cases.

\subsection{Case SI}
Let us consider the scalar sector in the case SI.  
The generating function is given by the linear combination of the first-class constraints, $G = \epsilon_i(t) \, {\cal C}_i^\beta$. Then ${\dot G}=\partial G/\partial t + \{G, H^S\}=0$ gives 
\begin{eqnarray}
\epsilon_1 + k^2 \epsilon_3 + \dot{\epsilon}_2  = 0, \qquad \epsilon_2+ \dot{\epsilon}_3 = 0\,.
\end{eqnarray}
Introducing the gauge parameter $\epsilon(t)$, we obtain the following relations, 
\begin{eqnarray}
\epsilon_1 = {\ddot \epsilon} - k^2 \epsilon, \qquad 
\epsilon_2 = -{\dot \epsilon}, \qquad 
\epsilon_3=\epsilon \,. 
\end{eqnarray}
Then the gauge transformation of the scalar components of $h_{\mu\nu}$ can be obtained from 
$\delta h_{\mu\nu} = \{h_{\mu\nu}, G\}$, 
\begin{eqnarray}
\delta \alpha = {k} {\dot \epsilon}, \qquad
\delta \beta = {\ddot \epsilon}-k^2 \epsilon, \qquad
\delta {\cal R} = 0, \qquad \delta {\cal E}= k {\dot \epsilon} \,.
\end{eqnarray}
One can easily check that the Lagrangian for the scalar mode is invariant under this gauge transformation. 	Taking into account the gauge transformation of the vector mode, 
	which is identical to the one in diffeomorphisms, the gauge transformation in the case SI can be covariantly written as 
	\begin{eqnarray}
	h_{\mu\nu}  \to h_{\mu\nu} + \partial_\mu \xi_\nu 
	+\partial_\nu \xi_\mu, \qquad 
	\textrm{with}~~\partial^\mu \xi_\mu = 0 
	\,.
	\label{TDiffg}
	\end{eqnarray}
	which is transverse diffeomorphisms~\cite{Alvarez:2006uu}.

To find the reduced Lagrangian, let us now define the gauge invariant variables,
\begin{eqnarray}
\widetilde{\alpha} =\alpha + {{\dot \beta}\over k} - {{\ddot {\cal E}} \over k^2}  \,, \qquad 
\widetilde{{\cal E}} =  {\cal E} - \alpha - 3{\cal R}\,.
\label{GIVI}
\end{eqnarray}
Note that the gauge invariant variable ${\widetilde {\cal E}}$ is nothing but the trace of $-h_{\mu\nu}$.
Using the gauge invariant variables introduced in the above, 
we rewrite the Lagrangian in terms of ${\widetilde \alpha}, {\widetilde {\cal E}},$ and ${\cal R}$ 
by eliminating $\beta$. And then after solving the constraint generated by the equation of motion for $\widetilde \alpha$, we can integrate out ${\cal R}$.  
Finally, we obtain the reduced action,  
\begin{eqnarray}
{\cal L}^S = \frac{4\kappa_1^2-4\kappa_1\kappa_3 + 3\kappa_3^2 + 8\kappa_1 \kappa_4}{2\kappa_1} 
\left[
{\dot {\widetilde {\cal E}}}{}^2 - \left(
k^2 + \frac{8\mu_2 \kappa_1 }{4\kappa_1^2-4\kappa_1\kappa_3 + 3\kappa_3^2 + 8\kappa_1 \kappa_4}
\right){\widetilde {\cal E}}^2
\right]\,.
\end{eqnarray}
Therefore, the conditions for avoiding ghost and tachyonic instabilities are 
\begin{eqnarray}
4\kappa_1^2-4\kappa_1\kappa_3 + 3\kappa_3^2 + 8\kappa_1 \kappa_4>0, \qquad \mu_2 >0 \,.
\end{eqnarray}

\subsection{Case SIIa}
In this case, the generating function is given by $G = \sum_{i,j}^2 \epsilon_{ij} {\cal C}_{ij}$, 
where $\epsilon_{ij}$ is the gauge parameter, and ${\cal C}_{11} = {\cal C}_1^\alpha, {\cal C}_{12}={\cal C}_2^\alpha, {\cal C}_{21}= {\cal C}_1^\beta,$ and ${\cal C}_{22}= {\cal C}_2^\beta$.
Then, the condition $\dot G =0$ gives two equations, 
\begin{eqnarray}
\epsilon_{11}  + \dot{\epsilon}_{12}+ \frac{2\kappa_1-\kappa_3}{4(\kappa_1-\kappa_3)} {k}\, \epsilon_{22}&=&0, \\
\epsilon_{21} + \dot{\epsilon}_{22}
-\frac{4(\kappa_1-\kappa_3)}{2\kappa_1-3\kappa_3}{k} \,\epsilon_{12} &=&0 \,.
\end{eqnarray}
These equations can be recasted into 
\begin{eqnarray}
\epsilon_{11} =  -\dot{\epsilon}_1
-\frac{2\kappa_1-\kappa_3}{4(\kappa_1-\kappa_3)} {k}\, \epsilon_{2} , \qquad
\epsilon_{12} = \epsilon_1, \qquad
\epsilon_{21} = \frac{4(\kappa_1-\kappa_3)}{2\kappa_1-3\kappa_3}{k}\, \epsilon_1 -\dot{\epsilon}_2, \qquad 
\epsilon_{22} = \epsilon_2 \,, 
\end{eqnarray}
where we have introduced two gauge parameters $\epsilon_1(t)$ and $\epsilon_2(t)$.
The gauge transformations are given by
\begin{align}
\delta \alpha &= -\dot{\epsilon}_1
-\frac{2\kappa_1-\kappa_3}{4(\kappa_1-\kappa_3)} {k}\, \epsilon_{2} \,, 
\qquad \qquad \qquad \qquad ~~~~
\delta \beta = \frac{4(\kappa_1-\kappa_3)}{2\kappa_1-3\kappa_3}{k}\, \epsilon_1 -\dot{\epsilon}_2 \,, \\
\delta {\cal R} &= (2 \kappa_1 - \kappa_3) \left[
\frac{\dot{\epsilon}_{1}}{2\kappa_1-3\kappa_3}  
-\frac{{k}\,\epsilon_{2}}{4(\kappa_1-\kappa_3)} \right] \,,
\qquad 
\delta {\cal E} = -{k}\, \epsilon_2 \,.
\end{align}
One can construct gauge invariant variables as follows, 
\begin{eqnarray}
 \alpha_\mathrm{IIa} =
\alpha  -\frac{2\kappa_1-\kappa_3}{4(\kappa_1-\kappa_3)} {\cal E} 
+
\frac{2\kappa_1-3\kappa_3}{4 (\kappa_1-\kappa_3)} \left( \frac{\dot{\beta}}{k} - \frac{\ddot{{\cal E}}}{k^2} \right) \,, \qquad 
{\cal E}_\mathrm{IIa} = {\cal E} - \alpha - \frac{2 \kappa_1 - 3 \kappa_3}{2 \kappa_1 - \kappa_3} \, {\cal R} \,.
\label{GIVIIb}
\end{eqnarray}
In the Einstein-Hilbert limit, $2\kappa_1 -\kappa_3=0$, ${\cal R}$ itself is gauge invariant just as in general relativity, and $ \alpha_\mathrm{IIa}$ coincides with $\widetilde{\alpha}$ defined in the case SI and also SIIb.
Using these gauge invariant variables, one can show that 
the reduced Lagrangian becomes zero as expected. 

Taking into account the gauge transformation of the vector mode, the gauge transformation in the case SIIa can be covariantly written as 
\begin{eqnarray}
h_{\mu\nu}  \to h_{\mu\nu} + \partial_\mu \xi_\nu 
+\partial_\nu \xi_\mu 
+b\, \partial_\rho \xi^\rho \eta_{\mu\nu} \,, 
\end{eqnarray}
where the constant $b$ is given by
\begin{eqnarray}
b= - \frac{2\kappa_1- \kappa_3}{2(\kappa_1-\kappa_3)} \,,
\end{eqnarray}
Note that the transformation in the case $b=0$, equivalently $2 \kappa_1 = \kappa_3$, reduce to
the standard gauge transformation in linearized general relativity.

\subsection{Case SIIb}
In the case SIIb, we first arrange the first class and second-class constraints. 
To do so, we define the following 
Hamiltonian including only the second-class constraint ${\cal C}^\alpha_1$, 
\begin{eqnarray}
H_T^S = H^S + \lambda_\alpha {\cal C}^\alpha_1\,.
\end{eqnarray}
Then the first-class constraints 
can be written as 
\begin{eqnarray}
{\widetilde {\cal C}}_2^\beta =
\{{\cal C}_1^\beta, H_T^S\}, \qquad 
{\widetilde {\cal C}}_3^\beta=
\{{\widetilde {\cal C}}_2^\beta , H_T^S\}, \quad 
\dot{\widetilde {\cal C}}{}_3^\beta =
\{{\widetilde {\cal C}}_3^\beta , H_T^S\}
= k^2 {\widetilde {\cal C}}_2^\beta \,.
\end{eqnarray}
Then as in the previous case, we define the generating function as $G = \epsilon_i(t) \, {\cal C}_i^\beta$, and ${\dot G}=\partial G/\partial t + \{G, H^S\}=0$ yields 
\begin{eqnarray}
\epsilon_1 + k^2 \epsilon_3 + \dot{\epsilon}_2  = 0, \qquad \epsilon_2+ \dot{\epsilon}_3 = 0\,.
\end{eqnarray}
In this case, one obtains the same equations as in the case SI. Thus the gauge transformations of $\alpha, \beta, {\cal R},$ and ${\cal E}$ are also the same as well as the gauge transformation in a covariant form, which is given by \eqref{TDiffg}.
Using the gauge invariant variables introduced in the case SI, 
one can show that the reduced Lagrangian after solving all the constraints becomes zero, implying no scalar degree of freedom.

\subsection{Case SIIc}
In this case, all the constraints are second-class, which implies that there is no gauge degree of freedom.
After integrating out the variables $\alpha$, $\beta$, and ${\cal R}$, we finally obtain the reduced Lagrangian for ${\cal E}$:
\begin{eqnarray}
{\cal L}^S = \frac{24 \mu_1^2}{(2 \kappa_1 k^2 + 3 \mu_1)^2}
\left[
\kappa_1 \dot{{\cal E}}^2 - (\kappa_1 k^2+ \mu_1) {{\cal E}}^2
\right]\,.
\end{eqnarray}
One can immediately see that the remaining scalar degree of freedom is always ghost-free as long as 
$\kappa_1 > 0$, and the conditions for avoiding tachyonic instability is given by $\mu_1>0$.


%

\end{document}